\documentclass[aps, longbibliography, superscriptaddress]{revtex4-2}
\usepackage{amsmath}
\usepackage{graphicx, textcomp}
\usepackage{enumerate,subeqnarray}
\usepackage{verbatim,natbib}
\usepackage{amsmath, amsthm, amssymb}
\usepackage{amsfonts}
\usepackage{subcaption, caption}
\usepackage[colorlinks=true, citecolor=red, linkcolor=blue, urlcolor=blue ]{hyperref}
\usepackage{booktabs, multirow}

\usepackage{orcidlink, etoolbox}

\begin{document}

\title{Modal analysis and optimisation of swimming active filaments}

\author{
John Severn\,\orcidlink{0009-0001-2098-4997} and Eric Lauga\,\orcidlink{0000-0002-8916-2545}}

\address{Department of Applied Mathematics and Theoretical Physics, University of Cambridge, Cambridge CB3 0WA, UK}

\begin{abstract}

Active flexible filaments form the classical continuum framework for modelling the locomotion of spermatozoa and algae driven by the periodic oscillation of flagella. This framework also applies to the locomotion of various artificial swimmers. Classical studies have quantified the relationship between internal forcing (localised or distributed internal moments or forces) and external output (filament shape and swimming speed). In this paper, we pose locomotion as a mathematical optimisation problem and demonstrate that the swimming of an isolated active filament can be accurately described and optimised using a small number of eigenmodes, significantly reducing computational complexity. In particular, we reveal that the motion of a filament with monophasic forcing, relevant to recently proposed artificial swimmers, is governed by exactly four forcing eigenmodes, only two of which are independent. We further present optimisations of such swimmers under various constraints.

\end{abstract}
\maketitle


\section{Introduction}

Microorganisms employ a variety of mechanisms in order to self-propel through viscous fluids~\cite{Lauga_Powers_2009, Berg_2003, Goldstein_2015, Gaffney_Gadêlha_Smith_Blake_Kirkman-Brown_2011, Wiggins_Goldstein_1998, Lauga_Goldstein_2012}. Due to their small size, their swimming is governed by low Reynolds number hydrodynamics, i.e.~Stokes flows~\cite{Leal_2010}{, which is the assumed flow regime for the swimmers (whether micro-scale, millimeter-scale, or larger) discussed in this paper}. In this regime, inertial effects in the fluid become negligible relative to viscous effects, and the locomotion kinematics are constrained by the Scallop Theorem:  time-reversible motion of the swimmer's body cannot produce any net swimming~\cite{Purcell_1977} (so named because a small scallop that repeatedly opens and closes would not be able to swim in the Stokes flow limit). 
Consequently, microorganisms must employ non-time-reversible  actuation in order to undergo net locomotion. Such mechanisms include the rotation of rigid helical flagellar filaments of bacteria, such as those employed by the model organism \textit{Escherichia coli}, that rotate to push the swimmer forwards~\cite{Berg_2003}, or the waving dynamics of the two flexible flagella of the green algae {\it Chlamyomonas} that act like arms to pull the swimmer forwards~\cite{Goldstein_2015}. Perhaps the most commonly known method of propulsion is that of spermatozoa, which utilises an undulating flexible flagellum to transmit travelling waves that push fluid backwards and hence push the swimmer forwards~\cite{Gaffney_Gadêlha_Smith_Blake_Kirkman-Brown_2011, Wiggins_Goldstein_1998}. 
 
It has long been known that the actuation of the spermatozoa flagellum is not simply localised to the point of attachment with the cell body, but is instead continuously distributed along the entire flagellum length~\cite{Machin_1958, Rikmenspoel_1965, Brokaw_1970, Rikmenspoel_1966}.  
This is facilitated by the axoneme, the internal structure of the flagellum, containing molecular motors that power the relative sliding of microtubules~\cite{Fawcett_1975, Inaba_2011}, leading to an effective (and sophisticated) distribution of internal  moments and forces that are  functions of both space (location along the flagellum) and time~\cite{Gaffney_Gadêlha_Smith_Blake_Kirkman-Brown_2011}.  Not only does understanding the swimming dynamics of spermatozoa have fundamental interest for cellular  biology and  fertility science~\cite{Fauci_Dillon_2006}, but the simple form of the spermatozoa model (a single flexible filament attached to a passive body) makes it an ideal basis for  theoretical studies in mathematical biology~\cite{Gray_Hancock_1955, Lauga_2007, Yu_Lauga_Hosoi_2006, Wiggins_Goldstein_1998, Camalet_Jülicher_Prost_1999, Spagnolie_Lauga_2010, Lauga_Eloy_2013}, and for the design and fabrication of experimental, artificial swimmers~\cite{Elgeti_Winkler_Gompper_2015, Fu_Wei_Yin_Yao_Wang_2021}, including those with biomedical applications \cite{Bunea_Taboryski_2020}.

 {{Whilst we aim to make the analysis in this paper as general as possible, we keep in mind two intuitive and motivational examples of such sperm-like swimmers.} Dreyfus et al.~\cite{Dreyfus_et_al_2005} constructed a micro-swimmer with a flagellum made of connected magnetic beads, actuated by an oscillating external magnetic field, that propelled a payload in the form of a red blood cell. This design was refined in later works to produce increasingly effective and diverse swimmers~\cite{Khalil_Dijkslag_Abelmann_Misra_2014, Khalil_Tabak_Hamed_Mitwally_Tawakol_Klingner_Sitti_2017, Hamilton_Gilbert_Petrov_Ogrin_2018, Jang_2015}. However, such swimmers are not truly self-propelled, relying on external magnetic fields. In a creative solution to this limitation, Williams et al.~\cite{Williams_Anand_Rajagopalan_Saif_2014} created a polymeric sperm-like {millimeter-scale} swimmer that was powered by heart muscle cells (cardiomyocytes) cultured onto the side of the flagellum, and this concept of muscle-powered swimmers has since then been diversely explored \cite{Holley_Nagarajan_Danielson_Zorlutuna_Park_2016, Cvetkovic_Raman_Chan_Williams_Tolish_Bajaj_Sakar_Asada_Saif_Bashir_2014, Yuan_Guo_Jin_Zhang_Yang_2023}. {We will later see that the magnetic swimmer of Ref.~\cite{Dreyfus_et_al_2005} and the biohybrid swimmer of Ref.~\cite{Williams_Anand_Rajagopalan_Saif_2014} provide key and intuitive examples for the application of the models developed in this paper.}}

From a fundamental physical point of view, flexible flagella are subject to three principal  forces: viscous drag from the surrounding fluid, active internal forcing from the biological activity at the axoneme level  {(or artificial equivalents)}, and passive elasticity of the flagella.  Due to the lack of inertia, these forces must always instantaneously balance, producing classical partial differential equations (PDEs) which govern the dynamics of the filament~\cite{Camalet_Jülicher_2000}.  These equations, known as the elastohydrodynamic (EHD) equations, directly relate the time-varying shape of the flagella   to the active internal forcing, and have been used to model both real spermatozoa~\cite{Gaffney_Gadêlha_Smith_Blake_Kirkman-Brown_2011} and  {sperm-like artificial swimmers~\cite{Elgeti_Winkler_Gompper_2015, Fu_Wei_Yin_Yao_Wang_2021, Bunea_Taboryski_2020, Dreyfus_et_al_2005, Williams_Anand_Rajagopalan_Saif_2014}.}

Mathematically, for a given active forcing inside the flagellum, the EHD equations can be solved to determine the filament motion~\cite{Gaffney_Gadêlha_Smith_Blake_Kirkman-Brown_2011}.  Then, since the total hydrodynamic forces and moments exerted by a swimming microorganisms must be zero at all times, the filament motion can be used to determine the swimming kinematics via a global force and moment balance.  These calculations are generally numerical, however a fully analytical approach becomes available if perturbations are assumed to be small relative to a straight flagellum.  However, to the authors' knowledge, no single formula for the swimming speed directly in terms of the active forcing (without the need to explicitly calculate the filament motion) has yet been offered, even in this linearised limit. 
In addition to this linearisation, proposed experimental artificial swimmers often utilise forcing that is entirely in phase~\cite{Dreyfus_et_al_2005, Williams_Anand_Rajagopalan_Saif_2014}, simplifying the problem further. Despite these simplifications, optimising the configuration of the swimmer, involving parameters such as filament elasticity, fluid viscosity, and active force distribution, remains a largely brute-force computational  {(or   experimental) }task \cite{Williams_Anand_Rajagopalan_Saif_2014, Khalil_Dijkslag_Abelmann_Misra_2014, Khalil_Tabak_Hamed_Mitwally_Tawakol_Klingner_Sitti_2017, Jang_2015}.

In this paper, we consider an idealised version of a biological or artificial flagellum: an active  flexible filament waving in a Stokes flow under small-amplitude forcing and with no head or cell body. In the first part of our paper, revisiting classical work, we solve the classical EHD equations and derive a new formula {directly linking} the active forcing to the swimming speed via a symmetric swimming speed function, bypassing the need to explicitly calculate the filament motion.
 In the second part of our paper, we  pose an optimisation problem wherein we maximise the swimming speed subject to a fixed forcing magnitude, and show that the solutions to this optimisation problem are the eigenmodes of the swimming speed function, which form an orthonormal basis for all possible forcing functions. We demonstrate optimisation procedures to maximise the swimming speed of the filament subject to a variety of constraints, resulting in a reduced computational complexity compared to classical methods.  We pay particular attention to the case of monophasic forcing, relevant to artificial swimmers studied experimentally~\cite{Dreyfus_et_al_2005, Williams_Anand_Rajagopalan_Saif_2014}. By applying the optimisation procedure to such filaments, we remarkably find that only four of the eigenvalues are non-zero{, and analytically calculate these eigenvalues and their corresponding eigenfunctions}.  Two of these eigenmodes are simply reflections of the other two (their eigenvalues being negatives), and one eigenvalue dominates the other under optimal physical parameter conditions.  Swimming is therefore governed approximately by just a single eigenvalue and eigenmode pair. Finally, we demonstrate that analysis can be applied to this eigenmode pair to optimise swimming far more efficiently than brute-force computation. 


The paper is organised as follows. In \S\ref{S:2} we {obtain} the full dimensionless EHD equations for a general moment forcing function.  We then linearise these equations for a  small internal forcing, and solve the resultant forced hyperdiffusion equation {to determine} the filament shape {(Eqs.~\eqref{eq:Psi_expression}, \eqref{eq:y1_expression} and \eqref{eq:psi1_expression})} in terms of the Green's function $G$ {(Appendix \ref{appendix:G})}. Using this solution and global force balances, we identify the swimming speed of the filament {(Eq.~\eqref{eq:U_G_swim})} entirely in terms of the moment forcing function and a symmetric swimming speed function $G_{swim}$ {(Eq.~\eqref{eq:G_swim})} that is constructed from $G$, bypassing the explicit solution for the {filament} shape. In \S\ref{S:3} we next {consider} an eigenvalue/eigenfunction problem for $G_{swim}$ that can be solved numerically (for arbitrary forcing phases) to obtain an {orthonormal} eigenmode basis {(Eq.~\eqref{eq:eigen})} from which to construct the forcing function{, and evaluate the swimming speed (Eq.~\eqref{eq:sum}). We also construct an optimisation procedure, which shows that these eigenfunctions produce local minima or maxima for the swimming speed, subject to a fixed forcing magnitude constraint.}

We {next} demonstrate the advantages of this method {using two key examples. We first consider a travelling wave of forcing,} obtaining similar results to those observed in biological spermatozoa~\cite{Gaffney_Gadêlha_Smith_Blake_Kirkman-Brown_2011}{, and identifying the optimal forcing wavelength, and a range of near-optimal values for the dimensionless parameter $Sp$ that denotes the relative elastic properties of the filament (Fig.~\ref{Figure_2}A)}.  {We then consider the case of monophasic forcing, analytically solving for the four non-zero eigenvalues and eigenfunctions. Considering} potential applications to experimental, artificial swimmers~\cite{Dreyfus_et_al_2005, Williams_Anand_Rajagopalan_Saif_2014}, we demonstrate analyses that can be used to optimise the swimming speed, subject to a variety of physical constraints and limitations. {In particular, and somewhat counter-intuitively, we show that swimming with a fixed total forcing magnitude is optimised in the limit of single-point actuation (rather than distributed forcing) and produces a far greater swimming speed than even eigenfunction forcing under the same constraint.} We conclude with \S\ref{S:4} with a summary of the key results, and offering a discussion of potential extensions of our modal approach to more complex waving swimmers.

\section{Classical elastohydrodynamics of active filaments}
\label{S:2}

A great deal of classical work has been done in modelling the response of the filament shape to both proximal and internal forcing, and how the resultant motion induces driving forces and locomotion~\cite{Wiggins_Goldstein_1998, Camalet_Jülicher_2000, Yu_Lauga_Hosoi_2006, Lauga_Powers_2009, Gaffney_Gadêlha_Smith_Blake_Kirkman-Brown_2011}. Here we summarise this  work, in particular presenting the classical linearised, dimensionless elastohydrodynamic (EHD) equations that balance the hydrodynamic, passive elastic, and active internal forces.  We then solve these for general active forcing using a Green's function, and thence derive the swimming speed function $G_{swim}$.  
Unless stated otherwise, we work below in the lab frame 
with standard Cartesian $(x,y)$ axes; this is the frame in which the fluid is stationary in the far field, and in which the filament achieves net displacement through swimming, as would be observed under a stationary microscope.

\subsection{Summary of classical work} 

\begin{figure}
\centering
\includegraphics[width = 0.8\linewidth]{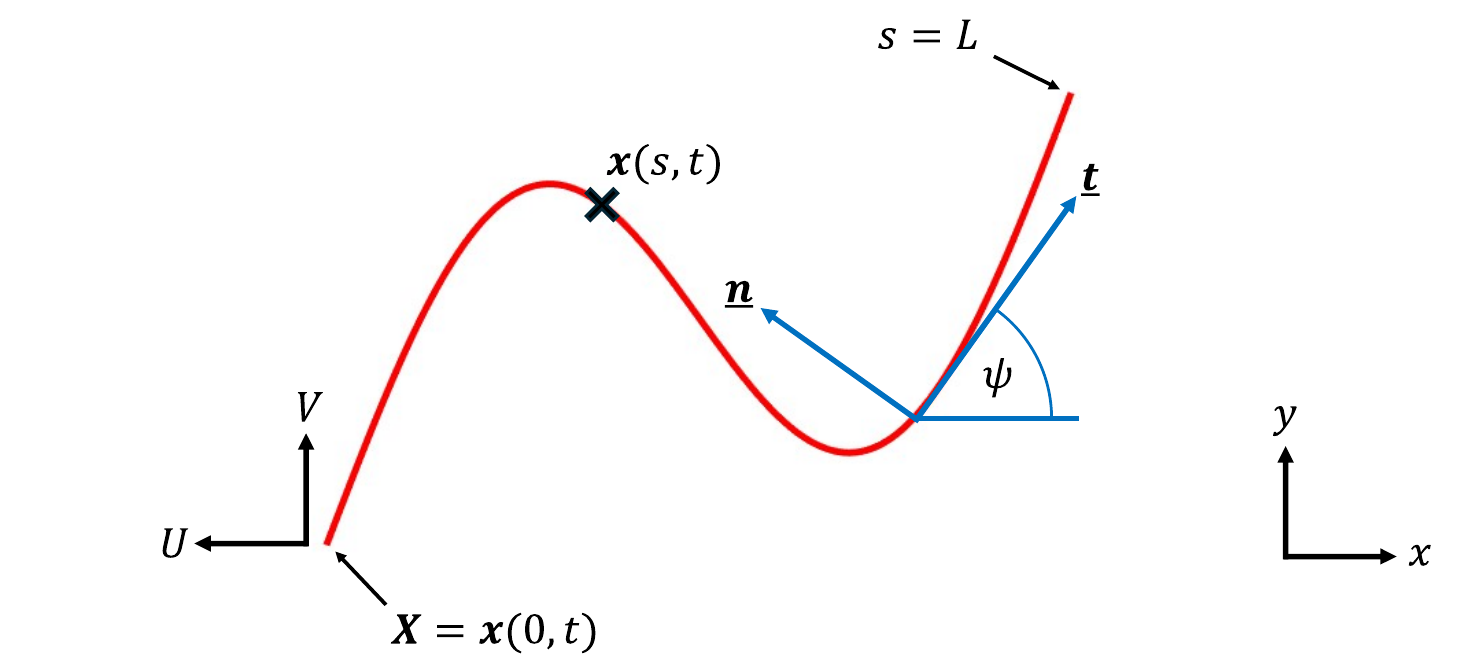}
\caption{{\bf Active filament.} Parameterisation of a (dimensional) filament of total length $L$. Notation includes: arc length  $0 \le s \le L$, tangent angle $\psi$, tangent vector $\mathbf{t}$, normal vector $\mathbf{n}$, tip position $\mathbf{X}$ and instantaneous tip velocity $\left(-U, V\right)$.}
\label{Figure_1}
\end{figure}

 {\bf{Parameterisation and notation.}} We begin by parameterising the filament of length $L$ by its arc length $0 \le s \le L$ and tangent angle $\psi(s, t)$, giving tangent vector $\mathbf{t} = \left(\cos\psi, \sin\psi\right)$ and normal vector ${\mathbf{n}} = \left(-\sin\psi, \cos\psi\right)$, as shown in Fig.~\ref{Figure_1}. The front tip, $s = 0$, has position $\mathbf{X}(t)$ and instantaneous velocity $\left(-U, V\right)$ (note the sign convention applied to $U$; forwards swimming gives $U > 0$). The position $\mathbf{x}(s, t)$ and velocity $\mathbf{u}(s, t)$ of a material point along the filament are then given by
\begin{align}
\mathbf{x}(s, t) &= \mathbf{X}(t) + \int_0^s \left(\begin{matrix}\cos\psi(s', t) \\ \sin\psi(s', t)\end{matrix}\right)~ds' \label{eq:x}, \\
\mathbf{u}(s, t) &= \left(\begin{matrix}-U(t) \\ V(t) \end{matrix}\right) + \int_0^s \left(\begin{matrix}-\sin\psi(s', t) \\ \cos\psi(s', t)\end{matrix}\right)\psi_t(s', t)~ds' \label{eq:u}.
\end{align}

 {\bf{{Dimensional hydrodynamic and} elastohydrodynamic (EHD) equations.}} The filament is subject to elastic forces with bending modulus $A${; as is standard,  $A$ has units Nm$^2$, and is a measure of the stress required to bend the filament, see Eq.~\eqref{eq:psi_dim}. The filament is also subject to} an active (internal) moment forcing $m(s,t)$, interpreted physically as the moment exerted by material at $s^+$ on material at $s^-$. {Therefore, $m_s$ is both the active moment per unit length acting on the filament, and the force exerted by material at $s^+$ on material at $s^-$. It follows that $m_{ss}$ can be interpreted as the active force per unit length acting on the filament. }{Fluid drag acting on the filament is calculated using resistive force theory~\cite{Lauga_book}, with parallel motion (i.e. motion in the direction of the long axis of the filament) incurring a drag force per unit length of $c_\|$ per unit filament speed, and perpendicular motion incurring a drag force per unit length of $c_\perp$ per unit filament speed, giving an overall hydrodynamic drag force per unit length
\begin{equation}
\mathbf{f} = -c_\perp\left(\mathbf{u}\cdot\mathbf{n}\right)\mathbf{n} - c_\|\left(\mathbf{u}\cdot\mathbf{t}\right)\mathbf{t} = -c_\perp\mathbf{u} + \left(c_\perp - c_\|\right)\left(\mathbf{u}\cdot\mathbf{t}\right)\mathbf{t}. \label{eq:f_dim}
\end{equation}

The motion of the filament is then determined by balancing the elastic, active and hydrodynamic forces.} For brevity, we omit the derivation of the classical EHD equations, often done through variational calculus~\cite{Camalet_Jülicher_2000}, and instead directly state  them:
{\begin{eqnarray}
\psi_t &=& \frac{1}{c_\perp}\left(-A\psi_{ssss} + m_{sss} + \psi_s\tau_s + \tau\psi_{ss}\right) + \frac{1}{c_\|}\psi_s\left(A\psi_s\psi_{ss} - \psi_sm_{s} +\tau_s\right), \label{eq:psi_dim}\\
\tau_{ss} - \frac{c_\|}{c_\perp}\psi_s^2\tau &=& \frac{c_\perp + c_\|}{c_\perp}\psi_s\left(m_{ss} - A\psi_{sss}\right) + \psi_{ss}\left(m_s - A\psi_{ss}\right).
\end{eqnarray}}
{Here $\tau$ is an elastic tension force (units of N) that enforces inextensibility. Note that the applied moment per unit length $m_s$ has replaced the internal moment per unit length $af$ in Ref. \cite{Camalet_Jülicher_2000}.} 

{\bf{Dimensionless equations.}} {We now apply non-dimensionalisation, scaling lengths with the filament length $L$, time with $1/\omega$ (a relevant angular forcing frequency) and moments and forces with $A/L$ and $A/L^2$ respectively. Therefore the form of Eq.~\eqref{eq:u} for the filament velocity is unchanged,
\begin{equation}
\mathbf{u}(s, t) = \left(\begin{matrix}-U(t) \\ V(t) \end{matrix}\right) + \int_0^s \left(\begin{matrix}-\sin\psi(s', t) \\ \cos\psi(s', t)\end{matrix}\right)\psi_t(s', t)~ds', \label{eq:u_dimensionless}
\end{equation}
though now $0 \le s \le 1$ and $\mathbf{u}$, $U$ and $V$ have (implicitly) been non-dimensionalised by scaling them with $L\omega$. Meanwhile, the dimensionless hydrodynamic drag force per unit length is given by
\begin{equation}
\mathbf{f} = Sp^4\left(-\mathbf{u} + \frac{c_\perp - c_\|}{c_\perp}\left(\mathbf{u}\cdot\mathbf{t}\right)\mathbf{t}\right), \label{eq:f_dimensionless}
\end{equation}
with the dimensionless `Sperm' number $ Sp$ defined as 
\begin{equation}
Sp = \frac{L}{\left(A/\omega c_\perp\right)^{1/4}} = \frac{L}{l_e},
\end{equation}
where $l_e$ is the elastic penetration length, interpreted physically as the typical length scale over which an elastically travelling displacement wave is damped by fluid drag. Finally,} we obtain the EHD equations  in dimensionless form as,
\begin{eqnarray}
Sp^4\psi_t &=& -\psi_{ssss} + \psi_s\tau_s + \psi_{ss}\tau + m_{sss} + \frac{c_\perp}{c_\|}\psi_s\left(\psi_s\psi_{ss} + \tau_s - m_s\psi_s\right)
\label{eq:psi},\\
\tau_{ss} - \frac{c_\|}{c_\perp}\psi_s^2\tau &=& \frac{c_\perp + c_\|}{c_\perp}\psi_s\left(-\psi_{sss} + m_{ss}\right) + \psi_{ss}\left(-\psi_{ss} + m_s\right).
\label{eq:tau}
\end{eqnarray}

 {\bf{Global force balance.}} By noting that the total hydrodynamic force must be zero in Stokes flow, we can integrate the dimensionless hydrodynamic force density, {Eq.~\eqref{eq:f_dimensionless},} along the length of the filament, {where $\mathbf{u}$ is given by Eq.~\eqref{eq:u_dimensionless},} giving a global dimensionless force balance as
\begin{equation}
\begin{split}
\left(\begin{matrix}0 \\ 0\end{matrix}\right) = &\int^1_0 \left\{ \left(\begin{matrix}(1 - \frac{c_\perp - c_\|}{c_\perp}\cos^2(\psi))U + \frac{c_\perp - c_\|}{c_\perp}\sin(\psi)\cos(\psi)V\\ (-1 + \frac{c_\perp - c_\|}{c_\perp}\sin^2(\psi))V - \frac{c_\perp - c_\|}{c_\perp}\sin(\psi)\cos(\psi)U \end{matrix}\right) \right. \\ &~~~~~+ \left. \left(\begin{matrix}\int^s_0\psi_t(s')\sin(\psi(s')) + \frac{c_\perp - c_\|}{c_\perp}\cos(\psi(s))\psi_t(s')(\sin(\psi(s) - \psi(s')))ds' \\ \int^s_0-\psi_t(s')\cos(\psi(s')) + \frac{c_\perp - c_\|}{c_\perp}\sin(\psi(s))\psi_t(s')(\sin(\psi(s) - \psi(s')))ds'\end{matrix}\right) \right\} ds ,
\end{split}
\label{eq:force}
\end{equation}
from which {(the dimensionless)} $U$ and $V$ will be determined. 

{\bf{Leading-order asymptotics.}} Next, we {assume} that the dimensionless forcing  is small, of the form $m = \epsilon m^{(1)}$, and look to solve the problem in powers of  $\epsilon \ll 1$. {Whilst the following analysis is therefore rigorously valid only for small $\epsilon$, previous results using similar analysis have demonstrated remarkable accuracy for even $\mathcal{O}(1)$ forcing~\cite{Yu_Lauga_Hosoi_2006}.} Note that changing $m$ to $-m$ would have no effect on $\tau(s, t)$ or $U(t)$, but would change the signs of $\psi(s, t)$ and $V(t)$. Furthermore, in the limit $\epsilon \to 0$, we must have $U \to 0$ and $\tau \to 0$. Therefore we deduce the following expansions in the small parameter $\epsilon$
\begin{equation}
\begin{split}
\psi = \epsilon\psi^{(1)} + \epsilon^3\psi^{(3)} + \ldots ~~~~~&~~~~~\tau = \epsilon^2\tau^{(2)} + \epsilon^4\tau^{(4)} + \ldots\\
U = \epsilon^2U^{(2)} + \epsilon^4U^{(4)} + \ldots ~~~~~&~~~~~ V = \epsilon V^{(1)} + \epsilon^3V^{(3)} + \ldots
\end{split}
\end{equation}
 {Note that the $\mathcal{O}(\epsilon^2)$ tension $\tau$ will, classically, prove absent from the leading-order problem and can henceforth be ignored.} We have also assumed that, in the limit $\epsilon \to 0$, the resting straight filament is aligned with the $x$ axis; $\psi = 0$. Linearising Eq.~\eqref{eq:psi} gives the classical hyperdiffusion equation for linear elastohydrodynamics
\begin{equation}
Sp^4\psi_{t}^{(1)} + \psi_{ssss}^{(1)} = m_{sss}^{(1)},
\label{eq:hyperdiffusion}
\end{equation}
which describes the linearised local force balance between hydrodynamic drag (first term), restoring elastic effects (second term) and active forcing (third term). In this linear limit, the dimensionless elastic restoring moment and force exerted by material at $s^+$ on material at $s^-$ are $-\psi^{(1)}_{s}$ and $-\psi^{(1)}_{ss}$ respectively.

{\bf{Boundary conditions.}}  {In the derivation of the full EHD equations through variational calculus in \cite{Camalet_Jülicher_2000}, boundary terms demand that both the force and torque provided by the filament itself (i.e.~excluding the effects of hydrodynamic drag) must be zero at both boundaries. In our dimensionless, linear system, this becomes the conditions
\begin{equation}
m^{(1)} - \psi^{(1)}_s = m^{(1)}_s - \psi^{(1)}_{ss} = 0.
\label{eq:classical_bcs}
\end{equation}
This is because said forces and torques are provided by material at $s^+$ acting on material at $s^-$, which cannot occur at the boundaries due to the lack of further material.~This provides sufficient boundary conditions for the problem, albeit dependent on the choice of $m$. However we will later show that these can be reduced to boundary conditions that are independent of $m$, by suitable integration of Eq.~\eqref{eq:hyperdiffusion}.}

{\bf{Leading-order swimming speed.}} Finally, {considering the global force balance Eq.~\eqref{eq:force} at leading order (specifically, $\mathcal{O}(\epsilon)$ in the $y$ component, and $\mathcal{O}(\epsilon^2)$ in the $x$ component)} yields leading order expressions for $V$ and $U$,
\begin{align}
V^{(1)} &= -\int^1_0 \left(\int^s_0 \psi_{t}^{(1)}(s') ds'\right) ds, \label{eq:V1}\\
U^{(2)} &= -\int^1_0 \frac{c_\perp - c_\|}{c_\|}\psi^{(1)}V^{(1)} + \left(\int^s_0 \frac{c_\perp}{c_\|}\psi_{t}^{(1)}(s')\psi^{(1)}(s') + \frac{c_\perp - c_\|}{c_\|}\psi_{t}^{(1)}(s')(\psi^{(1)}(s) - \psi^{(1)}(s')) ds'\right) ds. \label{eq:U2}
\end{align}
{We will also soon show that the leading-order torque balance is zero, as required for free swimming.}

\subsection{Integrated hyperdiffusion equation}
In preparation for solving Eq.~\eqref{eq:hyperdiffusion} for a general forcing function $m$ using a Green's function $G$, which will require setting $m^{(1)}$ to be a $\delta$-function, we will now integrate the equation three times. We define 
\begin{eqnarray}
y^{(1)}(s,t) &=& \frac{1}{Sp^4}\int^t_0 \left(m_{ss}^{(1)}(0, t') - \psi_{sss}^{(1)}(0, t')\right)~dt' + \int^s_0 \psi^{(1)}(s', t)~ds' ,\\
\beta(s,t) &=& \int^s_0 y^{(1)}(s', t)~ds',\\
\Psi(s,t) &= &\int^s_0 \beta(s', t)~ds'.
\end{eqnarray}
{Here, $y^{(1)}$ will soon be shown to be the leading-order vertical position of the filament as a function of $s$ and $t$;    $\beta$ and $\Psi$ have no discernible physical interpretation, besides their definitions in terms of $y^{(1)}$.} The governing equations for these are simply $Sp^4y_{t}^{(1)}+y_{ssss}^{(1)}=m_{ss}^{(1)}$ for $y^{(1)}$, and $Sp^4\beta_{t}~+~\beta_{ssss}=m_s^{(1)}$ for $\beta$, whilst the equation for $\Psi$ is
\begin{equation}\label{eq:216}
Sp^4\Psi_{t} + \Psi_{ssss} = m^{(1)}.
\end{equation}
 {We now see that the boundary conditions Eq.~\eqref{eq:classical_bcs}, which here are equivalent to $m^{(1)} - \Psi_{ssss} = m^{(1)}_s - \Psi_{sssss} = 0$, reduce to
\begin{equation}
\Psi_t = \Psi_{st} = 0,
\end{equation}
at both boundaries, and these are the boundary conditions we will henceforth consider.}

\subsection{Calculating $V^{(1)}$ and $U^{(2)}$ and verifying global torque balance}
We now obtain simple expressions for the velocity of the filament tip, for general time-periodic forcing $m^{(1)}$, by applying these governing equations and boundary conditions. Recalling Eq.~\eqref{eq:V1}, the vertical speed is given by
\begin{equation}
\begin{split}
V^{(1)} &= -\int^1_0 \left(\int^s_0 \psi_{t}^{(1)}(s') ds'\right) ds \\
&= -\frac{1}{Sp^4}\int^1_0\left(\int^s_0 m_{s's's'}^{(1)}(s') - \psi_{s's's's'}^{(1)}(s') ds'\right) ds \\
&= -\frac{1}{Sp^4}\int^1_0 \left(m_{ss}^{(1)}(s) - m_{ss}^{(1)}(0) - \psi_{sss}^{(1)}(s) + \psi_{sss}^{(1)}(0)\right) ds \\
&= \frac{1}{Sp^4}\left(m_{ss}^{(1)}(0) - \psi_{sss}^{(1)}(0)\right) \,= \, y_{t}^{(1)}(0, t) \, = \, \Psi_{sst}(0,t).
\end{split}
\end{equation}
Therefore, $y^{(1)}(0, t)$ can be understood as the vertical position of the front tip at time $t$ in this linearised limit. From the definition of $y^{(1)}$, we then see that $y^{(1)}(s, t)$ is the leading order vertical position of the point $s$ at time $t$. Finally, note that periodic forcing, i.e.~periodic $\Psi$, will therefore yield $\left<V^{(1)}\right> = 0$, where $\left<
\ldots \right>$ denotes the time-average over a period.

Similarly, we can determine the time-average of leading-order swimming speed, $U^{(2)}$. Noting that the time-average of $\psi^{(1)}\psi_{t}^{(1)}$ is zero for periodic forcing, we find
\begin{equation}
\begin{split}
\left<U^{(2)}\right> &= -\frac{c_\perp - c_\|}{c_\|}\int^1_0 \left\{ \left<\psi^{(1)}V^{(1)}\right> + \left(\int^s_0 \left<\psi_{t}^{(1)}(s')\psi^{(1)}(s)\right> ds'\right) \right\} ds \\
&= -\frac{c_\perp - c_\|}{c_\|}\int^1_0\left<\psi^{(1)}\left[V^{(1)} + \left(\int^s_0 \psi_{t}^{(1)}(s') ds'\right) \right]\right> ds\\
&= -\frac{c_\perp - c_\|}{c_\|}\int^1_0\left<\psi^{(1)}y_{t}^{(1)}\right> ds \,=\, -\frac{c_\perp - c_\|}{c_\|}\int^1_0\left<\Psi_{sss}\Psi_{sst}\right> ds.
\end{split}
\label{eq:U_classical}
\end{equation}
This is simply the classical propulsive force formula~\cite{Lauga_Powers_2009} divided by $c_\|$. Here it will be useful to define the reduced swimming speed as $\mathcal{U} = \frac{2c_\|}{c_\perp - c_\|}\left<U^{(2)}\right>$, giving
\begin{equation}\label{eq:222}
\mathcal{U} = -2\int^1_0\left<\Psi_{sss}\Psi_{sst}\right> ds.
\end{equation}

 {Finally, we can calculate the leading order global torque, measured about the tip, acting on the filament as
\begin{equation}
\mathcal{G}^{(1)} = \int_0^1 y^{(1)}_t s~ds = \int_0^1 \Psi_{sst}~s~ds.
\end{equation}
Recalling that $\Psi_{st} = \Psi_t = 0$ at both boundaries, we integrate by parts to obtain
\begin{equation}
\mathcal{G}^{(1)} = -\int_0^1 \Psi_{st} ~ds = 0,
\end{equation}
hence the leading order global torque $\mathcal{G}^{(1)}$ is indeed zero at all times.}

\subsection{Solving for filament motion using a Green's function}
We now consider the simple case of a moment forcing function $m^{(1)}(s, t) = \Re\left[f(s)e^{-i\phi(s)}e^{it}\right]$, where $f$ is real, $\phi(s)$ is the (real) phase function, and the angular frequency is $1$ thanks to non-dimensionalisation. The corresponding solution to Eq.~\eqref{eq:216} is given by  $\Psi = \Re\left[\Phi(s)e^{it}\right]$, where $\Phi$ is a complex function given by $Sp^4 i \Phi + \Phi_{ssss} = f(s)e^{-i\phi(s)}$, with solution
\begin{equation}
\Phi(s) = \int_0^1 G(s; \xi) f(\xi)e^{-i\phi(\xi)}~d\xi.
\end{equation}
$G(s;\xi)$ is the Green's function of the problem, given by 
\begin{equation}
Sp^4 i G + G'''' = \delta(s - \xi),
\end{equation} 
{and subject to $G = G_s = 0$ at both boundaries. }The full derivation and expression for $G$ are given in Appendix \ref{appendix:G}. In particular, $G$ is comprised entirely of the four natural modes $e^{ks}$, where $k^4 = -Sp^4 i$. Therefore, the solution $\Psi = \Re\left[\Phi(s)e^{it}\right]$ for general forcing function $m^{(1)}(s, t) = \Re\left[f(s)e^{-i\phi(s)}e^{it}\right]$ is written as
\begin{equation}
\Psi(s, t) = \Re\left[e^{it}\int_0^1 G(s; \xi)f(\xi)e^{-i\phi(\xi)}~d\xi\right]
\label{eq:Psi_expression}.
\end{equation}
{Recalling that $y^{(1)} = \Psi_{ss}$ and $\psi^{(1)} = \Psi_{sss}$, this then allows us to express the filament shape in terms of the derivatives of $G$ with respect to $s$,
\begin{equation}
y^{(1)}(s, t) = \Re\left[e^{it}\int_0^1 G''(s; \xi)f(\xi)e^{-i\phi(\xi)}~d\xi\right]
\label{eq:y1_expression},
\end{equation}
or equivalently,
\begin{equation}
\psi^{(1)}(s, t) = \Re\left[e^{it}\int_0^1 G'''(s; \xi)f(\xi)e^{-i\phi(\xi)}~d\xi\right]
\label{eq:psi1_expression}.
\end{equation}}
Note that $G$, and therefore $\Psi${, $y^{(1)}$ and $\psi^{(1)}$}, are dependent upon the dimensionless sperm number $Sp$ that parameterises the problem. We omit the explicit dependence on $Sp$ for brevity, but proceed with the understanding that any calculation is done for  a specific value of $Sp$. In particular, any optima that we identify are the optima for a specific value of $Sp$, and variation of $Sp$ will be necessary to determine global optima.

\subsection{Calculating the swimming speed using $G$}
The result in Eq.~\eqref{eq:222} can be used to calculate the swimming speed when the filament shape is known. We would first need to calculate $\Psi$ and its derivatives, perform a time-average, and finally an integral in $s$. However, it is possible to circumvent calculating the filament shape, and instead calculate the swimming speed directly from the forcing function and $G$, using a double integral.~For general moment forcing $m^{(1)}(s, t) = \Re\left[f(s)e^{-i\phi(s)}e^{it}\right]$, we  show in Appendix 
\ref{appendix:U}  that $\mathcal{U}$ is given by 
\begin{equation}
\mathcal{U} = -\int_{\xi_1 = 0}^1 \int_{\xi_2 = 0}^1 f(\xi_1)\Im\left[G'(\xi_1;\xi_2)e^{i(\phi(\xi_1) - \phi(\xi_2))}\right]f(\xi_2)~d\xi_2~d\xi_1
\label{eq:U_general}.
\end{equation}
{Importantly, the derivation of this equation in Appendix \ref{appendix:U} involves a time-average over a period. Therefore, if higher frequency modes are present within $m^{(1)}$ (i.e. $\omega = 2, 3, 4, \ldots$ under the current non-dimensionalisation, which is applied with regards to the fundamental mode) then their interactions will be averaged out and vanish since, for example $\left<\sin\left(at\right)\sin\left(bt\right)\right> = 0$ when $a$ and $b$ are distinct integers. Similarly, any constant forcing (i.e. $\omega = 0$) will have no effect on the swimming speed, since constant forcing cannot produce swimming, nor can any interactions between this constant forcing and non-zero frequency modes. Noting that each mode will have its own non-dimensionalisation, and therefore require its own unique re-dimensionalisation, we deduce that the overall dimensional swimming speed can be obtained simply as the sum of the dimensional swimming speeds corresponding to each individual temporal mode of $m^{(1)}$.} {Obviously, this equation for the swimming speed is not very intuitive due to the complex values of $G$ and $e^{i\phi}$. }We can further define real symmetric and antisymmetric functions as
\begin{align}
G_s(\xi_1, \xi_2) &= -\frac{1}{2}\left(\Im\left[G'(\xi_1; \xi_2)\right] + \Im\left[G'(\xi_2; \xi_1)\right]\right), \\
G_a(\xi_1, \xi_2) &= -\frac{1}{2}\left(\Re\left[G'(\xi_1; \xi_2)\right] - \Re\left[G'(\xi_2; \xi_1)\right]\right),
\end{align}
and these enable us to rewrite Eq.~\eqref{eq:U_general} as
\begin{equation}
\mathcal{U} = \int_{\xi_1 = 0}^1 \int_{\xi_2 = 0}^1 f(\xi_1)G_{swim}(\xi_1, \xi_2)f(\xi_2)~d\xi_2~d\xi_1,
\label{eq:U_G_swim}
\end{equation}
where
\begin{equation}
G_{swim}(\xi_1, \xi_2) = G_s(\xi_1, \xi_2)\cos\left(\phi(\xi_1) - \phi(\xi_2)\right) + G_a(\xi_1, \xi_2)\sin\left(\phi(\xi_1) - \phi(\xi_2)\right),
\label{eq:G_swim}
\end{equation}
{is the {(real)} swimming speed function. The result in Eq.~\eqref{eq:U_G_swim} is the first main result of this paper, showing the direct link between forcing (function $f$) and swimming (reduced speed $\mathcal{U}$). 

Once again, note that $G_{swim}$ is dependent upon $Sp$, and also on the phase function $\phi$, though we omit these dependences from the notation for brevity. By construction, $G_{swim}$ is real and symmetric for any phase function $\phi$. Once $G_s$ and $G_a$ are calculated, $\mathcal{U}$ can easily be numerically evaluated for any forcing magnitude $f(s)$ and phase $\phi(s)$.}

\section{Modal analysis}
\label{S:3}
By deriving and solving the classical forced hyperdiffusion equation for filament motion, we have determined Eq.~\eqref{eq:U_G_swim} for the swimming speed $\mathcal{U}$ of the filament entirely in terms of the forcing function $f$ and the real symmetric function $G_{swim}$ that depends on the sperm number $Sp$ and phase function $\phi$. 

There are however two practical concerns regarding Eq.~\eqref{eq:U_G_swim}. First, this equation has a quadratic computational complexity. If $G_{swim}$ has been computed, and $f$ is known, and the integrals are evaluated using some $N$-point numerical integration scheme (e.g.~trapezium approximation), then evaluating $\mathcal{U}$ will usually be an $\mathcal{O}(N^2)$ process. Second, if we are interested in maximising $\mathcal{U}$ over choice of $f$, there is no clear way to do this using Eq.~\eqref{eq:U_G_swim}. 

We now show that both concerns can be addressed by exploiting the real symmetric nature of $G_{swim}$ and deriving  a modal analysis of the system. We will show  that $\mathcal{U}$ can be well approximated using an $\mathcal{O}(N)$ or even $\mathcal{O}(1)$ process, and optimisation techniques for the choice of $f$ become available.

\subsection{Eigenfunctions and eigenvalues: theory}
The key point that allows to derive a modal analysis of this problem is to note that, as the continuum extension of a real symmetric matrix, $G_{swim}$ has an infinite basis of orthonormal eigenfunctions $g_n(\xi)$ and corresponding eigenvalues $\lambda_n$, dependent upon $Sp$ and $\phi$, and given by
\begin{equation}
\int_{0}^1 G_{swim}(\xi_1, \xi_2)g_n(\xi_2)~d\xi_2 = \lambda_n g_n(\xi_1) ,\qquad \int_0^1 g_m(\xi)g_n(\xi)~d\xi = \delta_{m n}.
\label{eq:eigen}
\end{equation}
We can express $f$ in terms of this basis, and hence express the swimming speed from Eq.~\eqref{eq:U_G_swim} via this modal approach, leading to
\begin{equation}
f(\xi) = \sum_{n = 1}^\infty a_ng_n(\xi), \qquad a_n = \int_0^1 f(\xi)g_n(\xi)~d\xi, \qquad \mathcal{U} = \sum_{n = 1}^\infty a_n^2\lambda_n.
\label{eq:sum}
\end{equation}
Since eigenvalues typically decay in magnitude as $n$ becomes large,   only a finite truncation of these series is usually necessary to produce accurate results. 

 \subsection{Eigenfunctions and eigenvalues: computation}
Algebraically calculating the eigenfunctions and eigenvalues of $G_{swim}$ is often technically possible by exploiting the definition of the Green's function as the solution to a differential equation (see Example B below). However it is usually easier to calculate them numerically by discretising $G_{swim}$ into an $N \times N$ real symmetric matrix and the eigenfunction $g_k$ into a real vector of length $N$,
\begin{equation}
G^{matrix}_{m n} = G_{swim}\left(\frac{2m-1}{2N}, \frac{2n-1}{2N}\right), \qquad g_{k, n}^{vector} = g_k\left(\frac{2n-1}{2N}\right),
\end{equation}
{where $1 \le m, n \le N$. }The discretised eigenfunctions (eigenvectors in this context) and corresponding eigenvalues of Eq.~\eqref{eq:eigen} are therefore given by approximating the integral numerically as
\begin{equation}
\mathbf{G}^{matrix}\mathbf{g}_k^{vector} \approx N\lambda_k \mathbf{g}_k^{vector},
\end{equation}
and so the eigenfunctions and eigenvalues can easily be calculated via standard methods of computing the eigenvectors and eigenvalues of a real symmetric matrix. Note that $N$ must be large enough to accurately capture the behaviour of $G_{swim}$ and its eigenfunctions; in practice we find that setting $N = 100$ comfortably achieves this, in the sense that increasing $N$ further, even to $N = 1000$, had no discernible effect on any of the results or figures discussed below. Setting $N = 100$ enables $\mathbf{G}^{matrix}$ to be evaluated, and its eigenfunctions and eigenvalues calculated, practically instantly on a laptop computer. We find that the eigenvalues typically decay quite quickly, and only a handful of eigenfunctions need be considered in practice. Importantly, the number of relevant eigenvalues does not noticeably change when increasing $N$, see Fig.~\ref{Figure_2}b below. Of course, the eigenfunctions and eigenvalues depend on the phase function $\phi$ and the sperm number $Sp$. Whilst $Sp$ is something that can be continuously varied to find an optimum, it is best to pre-determine $\phi$ based on physical or biological context, as we will soon demonstrate with two key examples.

\subsection{Using calculus of variations to prove the eigenfunctions produce optimal swimming}
To continue with this modal analysis, we first seek an intuitive understanding of the eigenfunctions and eigenvalues obtained above, in particular how they provide optimal choices for the forcing function. We first note that $f$ must be constrained in some way, otherwise we could simply, for example, double $f$ in order to quadruple $\mathcal{U}$. Various physical and biological constraints may be relevant, such as having a fixed rate of doing work (particularly relevant for biological cells such as spermatozoa) or engineering constraints that limit the choice of $f$ (relevant for artificial swimmers). 

From a mathematical perspective, the simplest constraint is one of fixed average magnitude,
\begin{equation}
\int_0^1 f(\xi)^2~d\xi = 1,
\label{eq:ffmc}
\end{equation}
which can be interpreted physically as the filament having a fixed total (squared) moment forcing. We then consider the variational optimisation of the swimming speed subject to this fixed forcing magnitude, yielding Lagrangian
\begin{equation}
\mathcal{L} = \int_{\xi_1 = 0}^1 \int_{\xi_2 = 0}^1 f(\xi_1)G_{swim}(\xi_1, \xi_2)f(\xi_2)~d\xi_2 d\xi_1 - \lambda\left[\int_0^1 f(\xi)^2~d\xi - 1\right],
\end{equation}
where $\lambda$ acts as the Lagrange multiplier of the variational problem, enforcing the constraint in Eq.~\eqref{eq:ffmc} (i.e.~the bracketed term is zero). Applying the standard perturbation $f \mapsto f + \delta f$ yields a linear change in the Lagrangian given by
\begin{equation}
\frac{1}{2}\delta \mathcal{L} = \int_{\xi_1 = 0}^1 \int_{\xi_2 = 0}^1 \delta f(\xi_1)G_{swim}(\xi_1, \xi_2)f(\xi_2)~d\xi_2 d\xi_1 - \lambda\int_0^1 \delta f(\xi)f(\xi)~d\xi.
\end{equation}
By writing this as
\begin{equation}
\frac{1}{2}\delta \mathcal{L} = \int_{\xi_1 = 0}^1 \delta f(\xi_1) \left[\int_{\xi_2 = 0}^1 G_{swim}(\xi_1, \xi_2)f(\xi_2)~ d\xi_2 -  \lambda f(\xi_1)\right]~d\xi_1,
\end{equation}
we establish the solution to this variational problem as the function $f$ (obeying the fixed forcing magnitude constraint) satisfying
\begin{equation}\label{eq:39}
\int_{\xi_2 = 0}^1 G_{swim}(\xi_1, \xi_2)f(\xi_2)~ d\xi_2 = \lambda f(\xi_1).
\end{equation}

The result in Eq.~\eqref{eq:39} shows that  the eigenfunctions $f$ of $G_{swim}$ are precisely the choice of forcing function which provide (local) optima to this variational problem. In other words, setting $f$ to be an eigenfunction of $G_{swim}$ necessarily yields a local maximum (or minimum) for the swimming speed. The global maximum (for each particular value of $Sp$ and $\phi$) can then be identified as the eigenfunction with the largest (positive) eigenvalue, which can then be maximised over all acceptable choices of $Sp$ and $\phi$ to obtain the true maximum swimming speed. Alternatively, swimming in the opposite direction can be maximised by choosing the eigenfunction with the largest (in magnitude) negative eigenvalue. It should be noted that this simple choice of $f$ is only optimal when the constraint upon $f$ truly is a fixed magnitude constraint; a  variety of other constraints may be relevant, such as a fixed rate of doing work, or constraints upon $f$ itself (see Example B below). In such cases, the full suite of eigenfunctions will have to be considered in general, though typically all but a handful of these can be neglected by virtue of small eigenvalues.


\subsection{Example A: Travelling wave forcing}
We now consider the application of our modal approach to two relevant examples. 
A biologically relevant situation is that of a travelling wave of forcing~\cite{Gaffney_Gadêlha_Smith_Blake_Kirkman-Brown_2011}. If we set $\phi(s) = 2\pi ks$ for some constant $k$, the overall moment forcing is then given by
\begin{equation}\label{eq:310}
m^{(1)}(s, t) = \Re\left[f(s)e^{i(t - 2\pi ks)}\right].
\end{equation}
This corresponds to a forcing wave of wavelength $1/k$ (i.e.~there are $k$ wavelengths on the flagellum) travelling at speed $1/2\pi k$ {in the backwards direction, i.e.~from the proximal end ($s = 0$) to the distal end ($s = 1$)}. This is approximately observed in real spermatozoa, both in high-viscosity, viscoelastic cervical mucus substitute ($k \approx 2-3$, $Sp \approx 24$) and low-viscosity \textit{in vitro} fertilisation medium with viscosity similar to that of water ($k \approx 1.5$, $Sp \approx 4$)~\cite{Smith_Gaffney_Gadêlha_Kapur_Kirkman‐Brown_2009, Gadêlha_Gaffney_Smith_Kirkman-Brown_2010, Gaffney_Gadêlha_Smith_Blake_Kirkman-Brown_2011}. 

With the moment forcing in Eq.~\eqref{eq:310}, the swimming speed function $G_{swim}$ is given by
\begin{equation}\label{eq:311}
G_{swim}(\xi_1, \xi_2) = G_s(\xi_1, \xi_2)\cos\left(2\pi k\left(\xi_1 - \xi_2\right)\right) + G_a(\xi_1, \xi_2)\sin\left(2\pi k\left(\xi_1 - \xi_2\right)\right).
\end{equation}
Note that the forcing strength $f$ is stationary and does not move with the wave. {One naturally expects swimming to occur in the direction opposite to the direction of wave propogation, i.e.~$\mathcal{U}k > 0$ under the current sign convention, and this is observed in spermatozoa~\cite{Smith_Gaffney_Gadêlha_Kapur_Kirkman‐Brown_2009, Gadêlha_Gaffney_Smith_Kirkman-Brown_2010, Gaffney_Gadêlha_Smith_Blake_Kirkman-Brown_2011}, with backwards travelling waves ($k > 0$) generating forwards swimming ($\mathcal{U} > 0$)}. Meanwhile, $k < 0$ corresponds to {forwards travelling waves}, which are generally not biologically observed.

\begin{figure}
\centering
\includegraphics[width = \linewidth]{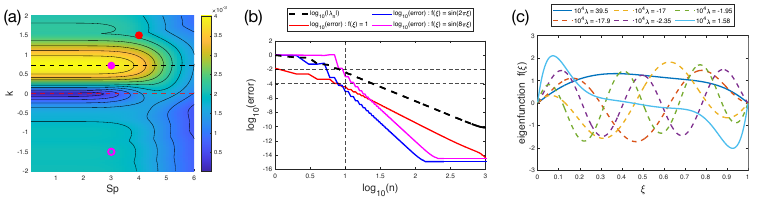}
\caption{{\bf Modal approach under travelling wave forcing.} (\textbf{a}) Heatmap of the largest positive eigenvalue for various values of $Sp$ and $k$, with $k = 0$ (dashed red line) and the optimal $k \approx 0.72$ (dashed black line) indicated. Examples of spermatozoa swimming through \textit{in vitro} fertilisation medium (red, $k = 1.5$, $Sp = 4$~\cite{Smith_Gaffney_Gadêlha_Kapur_Kirkman‐Brown_2009, Gadêlha_Gaffney_Smith_Kirkman-Brown_2010, Gaffney_Gadêlha_Smith_Blake_Kirkman-Brown_2011}) and a near-optimal active filament (solid pink, $k = 0.72$, $Sp = 3$) are also shown, as well as optimal swimming in the same direction as wave propagation (hollow pink, $k = -1.5$, $Sp = 3$). (\textbf{b}) Log-log plot (base $10$)  showing the errors incurred as the truncation of Eq.~\eqref{eq:sum} is varied, for a near-optimal filament ($k = 0.72$, $Sp = 3$). Eigenvalues are arranged in order of decreasing magnitude, normalised relative to the largest/first eigenvalue, and their decay as $n$ increases is given (thick dashed black line). Also plotted is the relative error incurred by truncating Eq.~\eqref{eq:sum} at the n-th term, normalised relative to the exact swimming speed calculated using Eq.~\eqref{eq:U_G_swim}, for simple forcing functions $f(\xi) \equiv 1$ (red solid line), $f(\xi) = \sin\left(2\pi\xi\right)$ (blue) and $f(\xi) = \sin\left(8\pi\xi\right)$ (pink). (\textbf{c}) The eigenfunctions for the six eigenvalues with largest magnitudes, {normalised using the fixed forcing magnitude constraint, Eq.~\eqref{eq:ffmc}}. The first and sixth eigenvalues are positive (solid lines) whilst the others are negative (dashed lines).}
\label{Figure_2}
\end{figure}

\subsubsection{Results of modal analysis}
\paragraph{Optimal swimming mode.}
For each $Sp$ and $k$, we next apply the above modal approach to compute the eigenvalues of Eq.~\eqref{eq:311}. The magnitude of largest positive eigenvalue  is plotted in  Fig.~\ref{Figure_2}a as a function of  $Sp$ and $k$ (allowing both signs for $k$). For each value of the parameters, this corresponds to the optimal swimming speed under the constraint of fixed forcing magnitude.   Note that, given the distributed forcing, swimming speed does not tend to zero as $Sp \to 0$, in contrast to classical results for single-point actuation~\cite{Lauga_2007}.  However, the swimming speed does tends to zero as $Sp$ becomes large, corresponding to elastic waves that decay over very short dimensionless length scales.

We observe that, {in accordance with the intuition that $\mathcal{U}k > 0$} in the biological world  {(i.e.~swimming occurs in the direction opposite to wave propagation)},   the optimal $k$ (i.e.~that which maximises forward swimming) is positive, around $k = 0.72$ (dashed  {black} line in Fig.~\ref{Figure_2}a).  In fact, there is a large region of near-optimality, for a range of values of both $k$ and $Sp$, and   the dependence on $Sp$ within this region is extremely weak. This means that there is a great deal of freedom in choosing $k$ and especially $Sp$ to optimise the forward swimming speed.  {Note that the the actual optimum has $Sp \to 0$, which is obviously physically impossible. Therefore we often select a "near-optimum'' (solid pink dot in Fig.~\ref{Figure_2}a) that could correspond to a real swimmer, without any notable sacrifice to swimming speed.}

 {Remarkably, the conditions for near-optimal  swimming of this isolated filament (solid pink dot in Fig.~\ref{Figure_2}a) are similar to the conditions observed in real spermatozoa swimming through water-like \textit{in vitro} fertilisation medium (red dot in Fig.~\ref{Figure_2}a)~\cite{Smith_Gaffney_Gadêlha_Kapur_Kirkman‐Brown_2009, Gadêlha_Gaffney_Smith_Kirkman-Brown_2010, Gaffney_Gadêlha_Smith_Blake_Kirkman-Brown_2011}. This  suggests that despite the various assumptions made, such as the  {linearisation and} lack of a head, an active filament model reasonably approximates biological spermatozoa. 
 
 Interestingly, $\mathcal{U}k < 0$ is also possible, and surprisingly effective (local optimum in the region $\mathcal{U}k < 0$ shown by the hollow pink dot in Fig.~\ref{Figure_2}a), with forwards travelling waves able to generate around half the maximum speed compared to the $\mathcal{U}k > 0$ case.}

\paragraph{Superposition of  modes.}

{If $f$ cannot be freely chosen, or is not subject to a fixed forcing magnitude, we must use the full suite of eigenfunctions and eigenvalues. Fortunately, the eigenvalues decay rapidly, as shown in Fig.~\ref{Figure_2}b (thick dashed black line) for a near-optimal filament with $k = 0.72$ and $Sp = 3$. Of course, a small eigenvalue can still contribute significantly to $\mathcal{U}$ if the corresponding coefficient $a_n$ is large enough. For a typical forcing function $f(\xi) \equiv 1$, approximately observed in spermatozoa~\cite{Gaffney_Gadêlha_Smith_Blake_Kirkman-Brown_2011}, Fig.~\ref{Figure_2}b also shows that the relative errors in approximating $\mathcal{U}$ using a truncation of Eq.~\eqref{eq:sum}, compared to using the exact expression in Eq.~\eqref{eq:U_G_swim}, also decays rapidly, and only a few terms need be retained to obtain an accurate result. 

For comparison, we also show in Fig.~\ref{Figure_2}b   the corresponding results obtained by setting $f(\xi) = \sin\left(2\pi\xi\right)$ and $f(\xi) = \sin\left(8\pi\xi\right)$. Overall, only a small truncation is needed, with only ten (out of a thousand) terms being sufficient to obtain an error of less than $0.01\%$ for both $f(\xi) \equiv 1$ and $f(\xi) = \sin\left(2\pi\xi\right)$, with  the higher-frequency forcing $f(\xi) = \sin\left(8\pi\xi\right)$ incurring an error of around $0.1\%$ using the same number of modes.

This increase in error induced by higher frequency forcing can be understood physically by examining the form of the eigenfunctions. The six eigenvalues with largest magnitudes, and their corresponding eigenfunctions (normalised using the fixed forcing magnitude constraint, Eq.~\eqref{eq:ffmc}), are shown in Fig.~\ref{Figure_2}c, again for a near-optimal filament with $k = 0.72$ and $Sp = 3$. Despite the fact that the forcing wave is travelling {backwards} (i.e.~$k > 0$) four of these six eigenvalues are negative (the largest eigenvalue is positive however, and more than twice the magnitude of any other eigenvalue). Furthermore, the negative eigenvalues appear to have eigenfunctions which are significantly more oscillatory than their counterparts for positive eigenvalues, and the eigenfunction of the largest eigenvalue is the least oscillatory of all. In fact, the frequency of the eigenmodes increases as the magnitudes of the eigenvalues decrease, explaining the greater errors observed for higher frequency forcing seen in Fig.~\ref{Figure_2}b. Finally, note that the eigenfunction of the largest eigenvalue is close to the actual forcing observed in real spermatozoa~\cite{Gaffney_Gadêlha_Smith_Blake_Kirkman-Brown_2011}.}

It is important to note that these results do not discernibly change when increasing $N$, hence the number of relevant eigenfunctions and eigenvalues does not change as $N$ is further increased. We can therefore generally truncate Eq.~\eqref{eq:sum} to a fairly small number of modes, whilst still retaining excellent accuracy in calculating the swimming speed.

\subsubsection{Computational considerations}

When $f$ can be freely varied, and is subject to a fixed forcing magnitude, we immediately obtain the optimal $f$ for fixed $Sp$ and $k$ as the eigenfunction with the largest eigenvalue. Otherwise, we use a finite truncation of Eq.~\eqref{eq:sum}, and this provides multiple computational advantages compared to standard methods of calculating $\mathcal{U}$.

For any given $f$, the swimming speed could be evaluated without the use of the modal approach, using Eq.~\eqref{eq:U_G_swim}, by performing a double integral.~Assuming said integrals are calculated numerically, using some $N$-point scheme (such as a trapezium approximation), this represents a computational complexity of $\mathcal{O}(N^2)$. However, using the modal approach, with a finite truncation of the eigenfunctions in Eq.~\eqref{eq:sum} (e.g.~retaining the ten largest-in-magnitude eigenvalues), we instead only need to calculate a fixed number of single integrals, and so the overall computational complexity of calculating $\mathcal{U}$ is $\mathcal{O}(N)$, with a slight trade-off of accuracy in the result due to the neglected eigenfunctions. Of course, this requires us to know the eigenfunctions and eigenvalues, however once these are calculated, they can be applied to calculate $\mathcal{U}$ in $\mathcal{O}(N)$ time for any $f$.

This advantage is even more striking  if, instead of choosing $f$ and then calculating integrals to identify the coefficients $a_n$ of Eq.~\eqref{eq:sum}, we do the converse: select the $a_n$, and use these to calculate $f$. Not only does this fully circumvent the need to calculate the integrals, enabling $\mathcal{U}$ to be computed with $\mathcal{O}(1)$ complexity, but it reduces the phase space of $f$ from an infinite-dimensional one (where $f$ can be continuously varied) to one of finite dimensions. Then the optimal $f$, for that particular $Sp$ and $k$, can be identified easily by maximising $\mathcal{U}$ through variation of the $a_n$.

Clearly, the actual procedure to calculate $\mathcal{U}$ will depend on the constraints imposed on $f$. Furthermore, these methods do not provide any way to maximise $\mathcal{U}$ over all $Sp$ and $k$. 


 \subsection{Example B: Monophasic forcing}

Having considered an example that is biologically relevant, we now turn one that is applicable to the design of artificial swimmers. Due to the difficulties of engineering on such small scales, forcing is typically simple in form. Two notable examples of experimental artificial swimmers are the filament made of magnetic beads used in Ref.~\cite{Dreyfus_et_al_2005}, and the polymeric flagellum actuated by cardiomyocytes used in Ref.~\cite{Williams_Anand_Rajagopalan_Saif_2014}. In both of these cases, forcing generally acts in phase, $\phi \equiv 0$ (equivalent to the $k = 0$ case in the previous example){, with the former applying distributed forcing and the latter utilising point actuation.} When taking $\phi \equiv 0$, $G_{swim}$ is given simply by
\begin{equation}\label{eq:Gsimmono}
G_{swim}(\xi_1, \xi_2) = G_s(\xi_1, \xi_2) = -\frac{1}{2}\left(\Im\left[G'(\xi_1; \xi_2)\right] + \Im\left[G'(\xi_2; \xi_1)\right]\right).
\end{equation}

In this section, we apply our modal approach to develop optimisation techniques for such artificial swimmers. We remarkably find that only (exactly) four of the eigenvalues are non-zero. Two of these are simply negatives of the other two, their eigenfunctions being reflections, due to the symmetry of the filament. We start below by analytically proving that only four of the eigenvalues are non-zero, followed by developing an analytic method to calculate the corresponding eigenfunctions. We then apply these results to establish exact formulae for the swimming speed $\mathcal{U}$, before making various approximations in order to optimise the configuration of an artificial swimmer, and demonstrating this using simple examples.

\subsubsection{Only four eigenvalues are non-zero}
Given the swimming speed function in Eq.~\eqref{eq:Gsimmono}, the eigenfunction $f$ with eigenvalue $\lambda$ satisfies
\begin{equation}
\begin{split}
\lambda f(\xi_1) &= \int_0^1 G_{swim}(\xi_1, \xi_2)f(\xi_2)~d\xi_2 \\
&= \frac{1}{2}\Im\left[\int_0^1 -G'(\xi_1; \xi_2)f(\xi_2) - G'(\xi_2; \xi_1)f(\xi_2)~d\xi_2\right] \\
&= \frac{1}{2}\Im\left[-\frac{\partial}{\partial \xi_1}\left(\int_0^1 G(\xi_1; \xi_2)f(\xi_2)~d\xi_2\right) + \int_0^1 G(\xi_2; \xi_1)\frac{\partial}{\partial \xi_2}\left(f(\xi_2)\right)~d\xi_2\right] \\
&= \frac{1}{2}\Im\left[-I_1'(\xi_1) + I_2(\xi_1)\right] \\
&= \Re\left[I_3\right],
\end{split}
\label{eq:I3}
\end{equation}
where $I_1$ and $I_2$ are the solutions to the differential equations 
\begin{align}
Sp^4 i I_1(\xi_1) + I_1''''(\xi_1) &= f(\xi_1) \label{eq:I1}, \\
Sp^4 i I_2(\xi_1) + I_2''''(\xi_1) &= f'(\xi_1) \label{eq:I2},
\end{align}
and $I_3 = \frac{1}{2}i\left(I_1' - I_2\right)$ satisfies the equation
\begin{equation}
Sp^4 i I_3(\xi_1) + I_3''''(\xi_1) = 0,
\end{equation}
with $I(0) = I(1) = I'(0) = I'(1) = 0$ for {both $I_1$ and $I_2$}. 

{Defining, as in Appendix \ref{appendix:G}, $\eta = e^{-\pi i/8}$, }we therefore deduce that
\begin{equation}
I_3(\xi_1) = Ae^{Sp\eta \xi_1} + Be^{Sp\eta i \xi_1} + Ce^{-Sp\eta \xi_1} + De^{-Sp\eta i \xi_1},
\end{equation}
consists entirely of the natural modes of the filament. Therefore, since the eigenfunction $f$ must satisfy $\lambda f(\xi_1) = \Re\left[I_3\right]$, we deduce that $\lambda = 0$ whenever $f$ contains a non-natural mode, and so there are only finitely many eigenfunctions with non-zero eigenvalues, each constructed entirely using the natural modes. Using complex coefficients, $I_3$ has only four modes, whilst using real coefficients, $f$ has eight modes. Note however that $I_3(0) = I_3(1) = 0$, which reduces the number of independent modes of $I_3$ to two, therefore reducing the number of independent modes of $f$ to four. Hence there are at most four eigenfunctions $f$ with non-zero eigenvalue, and by symmetry, two of these eigenfunctions will simply be reflections of the other two, their eigenvalues being negatives. 

We can identify the $I_3$ corresponding to each mode of $f$, and equating modes then results in an eigenvector problem, allowing us to find the eigenfunctions $f$ and their corresponding eigenvalues. This can be done using a standard method (see Appendix \ref{appendix:eig}) to obtain an $8 \times 8$ system with four zero eigenvalues. Alternatively, as we now show, we can instead  incorporate the boundary conditions for $I_3$ directly into the eigenfunction calculation to obtain a simpler $4 \times 4$ system that fully identifies the eigenfunctions and eigenvalues, and demonstrates their symmetry.

\subsubsection{Analytic calculation of eigenvalues}
The function $I_3$ must obey $I_3(0) = I_3(1) = 0$, and therefore, by expressing two of the coefficients in terms of the other two, we find that, regardless of $f$, we can express $I_3$ in the form
\begin{equation}
I_3(\xi_1) = Af_s(\xi_1) + Bf_a(\xi_1),
\label{eq:I32}
\end{equation}
for symmetric {($f_s(1-\xi_1) = f_s(\xi_1)$)} and antisymmetric {($f_a(1-\xi_1) = -f_a(\xi_1)$)} functions
\begin{align}
f_s(\xi_1) &= \frac{e^{Sp\eta\xi_1} + e^{Sp\eta}e^{-Sp\eta \xi_1}}{1 + e^{Sp\eta}} - \frac{e^{Sp\eta i\xi_1} + e^{Sp\eta i}e^{-Sp\eta i \xi_1}}{1 + e^{Sp\eta i}}, \\
f_a(\xi_1) &= \frac{e^{Sp\eta\xi_1} - e^{Sp\eta}e^{-Sp\eta \xi_1}}{1 - e^{Sp\eta}} - \frac{e^{Sp\eta i\xi_1} - e^{Sp\eta i}e^{-Sp\eta i \xi_1}}{1 - e^{Sp\eta i}}.
\end{align}
The forcing $f$ must also have this form to be an eigenfunction with non-zero eigenvalue, and in particular we can calculate $I_1$ and $I_2$ for the four different modes of $f$. For $f(\xi) = \Re\left[bf_s(\xi_1)\right]$, where $b = 1$ or $b = i$, we have
\begin{equation}
\begin{split}
I_1^{(b, s)}(\xi_1) &= \frac{b\xi_1}{8Sp^3\eta^3}\left(\frac{e^{Sp\eta\xi_1} - e^{Sp\eta}e^{-Sp\eta\xi_1}}{1+e^{Sp\eta}} - i\frac{e^{Sp\eta i\xi_1} - e^{Sp\eta i}e^{-Sp\eta i\xi_1}}{1+e^{Sp\eta i}}\right) + \frac{b^*f_s(\xi_1)^*}{4Sp^4 i} \\
&+ A_1^{(b, s)}e^{Sp\eta\xi_1} + B_1^{(b, s)}e^{Sp\eta i\xi_1} + C_1^{(b, s)}e^{-Sp\eta\xi_1} + D_1^{(b, s)}e^{-Sp\eta i\xi_1}, \\
I_2^{(b, s)}(\xi_1) &= \frac{b\xi_1}{8Sp^2\eta^2}\left(\frac{e^{Sp\eta\xi_1} + e^{Sp\eta}e^{-Sp\eta\xi_1}}{1+e^{Sp\eta}} + \frac{e^{Sp\eta i\xi_1} + e^{Sp\eta i}e^{-Sp\eta i\xi_1}}{1+e^{Sp\eta i}}\right) + \frac{b^*f_s'(\xi_1)^*}{4Sp^4 i} \\
&+ A_2^{(b, s)}e^{Sp\eta\xi_1} + B_2^{(b, s)}e^{Sp\eta i\xi_1} + C_2^{(b, s)}e^{-Sp\eta\xi_1} + D_2^{(b, s)}e^{-Sp\eta i\xi_1}.
\end{split}
\end{equation}
Meanwhile, if $f(\xi) = \Re\left[bf_a(\xi_1)\right]$, where $b = 1$ or $b = i$, we obtain
\begin{equation}
\begin{split}
I_1^{(b, a)}(\xi_1) &= \frac{b\xi_1}{8Sp^3\eta^3}\left(\frac{e^{Sp\eta\xi_1} + e^{Sp\eta}e^{-Sp\eta\xi_1}}{1-e^{Sp\eta}} - i\frac{e^{Sp\eta i\xi_1} + e^{Sp\eta i}e^{-Sp\eta i\xi_1}}{1-e^{Sp\eta i}}\right) + \frac{b^*f_a(\xi_1)^*}{4Sp^4 i} \\
&+ A_1^{(b, a)}e^{Sp\eta\xi_1} + B_1^{(b, a)}e^{Sp\eta i\xi_1} + C_1^{(b, a)}e^{-Sp\eta\xi_1} + D_1^{(b, a)}e^{-Sp\eta i\xi_1} ,\\
I_2^{(b, a)}(\xi_1) &= \frac{b\xi_1}{8Sp^2\eta^2}\left(\frac{e^{Sp\eta\xi_1} - e^{Sp\eta}e^{-Sp\eta\xi_1}}{1-e^{Sp\eta}} + \frac{e^{Sp\eta i\xi_1} - e^{Sp\eta i}e^{-Sp\eta i\xi_1}}{1-e^{Sp\eta i}}\right) + \frac{b^*f_a'(\xi_1)^*}{4Sp^4 i} \\
&+ A_2^{(b, a)}e^{Sp\eta\xi_1} + B_2^{(b, a)}e^{Sp\eta i\xi_1} + C_2^{(b, a)}e^{-Sp\eta\xi_1} + D_2^{(b, a)}e^{-Sp\eta i\xi_1}.
\end{split}
\end{equation}
Here there are 32 coefficients that must be evaluated by applying the boundary conditions on $I_1$ and $I_2$. According to Eqs.~\eqref{eq:I1} and \eqref{eq:I2}, and the definition of $I_3$, we see that a symmetric $f$ results in an antisymmetric $I_3$, and an antisymmetric $f$ results in a symmetric $I_3$. Recalling the form that $I_3$ must take, Eq.~\eqref{eq:I32}, and that the non-natural modes of $I_1$ and $I_2$ vanish in $I_3$, we deduce that
\begin{align}
I_3^{(b, s)}(\xi_1) = \frac{1}{2}i\left(\left(I_1^{(b, s)}\right)' - I_2^{(b, s)}\right) &= E_3^{(b, s)}f_a(\xi_1), \\
I_3^{(b, a)}(\xi_1) = \frac{1}{2}i\left(\left(I_1^{(b, a)}\right)' - I_2^{(b, a)}\right) &= E_3^{(b, a)}f_s(\xi_1),
\end{align}
where each of the new coefficients is given by
\begin{align}
E_3^{(b, a)} &= \frac{1}{2}i\left(1 + e^{Sp\eta}\right)\left(Sp\eta A_1^{(b, a)} - A_2^{(b, a)}\right), \\
E_3^{(b, s)} &= \frac{1}{2}i\left(1 - e^{Sp\eta}\right)\left(Sp\eta A_1^{(b, s)} - A_2^{(b, s)}\right).
\end{align}
These four coefficients can be calculated for any given $Sp$, and this gives us a fully determined system. If we write the eigenfunction as
\begin{equation}
f(\xi_1) = A_f\Re\left[f_s(\xi_1)\right] + B_f\Re\left[f_a(\xi_1)\right] + C_f\Re\left[if_s(\xi_1)\right] + D_f\Re\left[if_a(\xi_1)\right],
\end{equation}
then this results in an $I_3$ given by
\begin{equation}
I_3(\xi_1) = \left(A_f E_3^{(1, s)} + C_f E_3^{(i, s)}\right)f_a(\xi_1) + \left(B_f E_3^{(1, a)} + D_f E_3^{(i, a)}\right)f_s(\xi_1).
\end{equation}
Recalling that the coefficients of $f$ are real, and that $\lambda f = \Re\left[I_3\right]$, we finally obtain  an eigenvector problem
\begin{equation}
\lambda\left[\begin{matrix}A_f \\ B_f \\ C_f \\ D_f \end{matrix}\right] = \left[\begin{matrix}0 & \Re\left[E_3^{(1, a)}\right] & 0 & \Re\left[E_3^{(i, a)}\right] \\ \Re\left[E_3^{(1, s)}\right] & 0 & \Re\left[E_3^{(i, s)}\right] & 0 \\ 0 & \Im\left[E_3^{(1, a)}\right] & 0 & \Im\left[E_3^{(i, a)}\right] \\ \Im\left[E_3^{(1, s)}\right] & 0 & \Im\left[E_3^{(i, s)}\right] & 0\end{matrix}\right]\left[\begin{matrix}A_f \\ B_f \\ C_f \\ D_f \end{matrix}\right] .
\end{equation}
This equation can easily be solved for the coefficients of $f$, therefore identifying the eigenfunctions and eigenvalues. Furthermore, we see that any solution has the property that a change $B_f \mapsto -B_f, D_f \mapsto -D_f, \lambda \mapsto -\lambda$ still satisfies this equation, which is equivalent to a change $f(\xi) \mapsto f(1-\xi)$ with negative eigenvalue, as required by symmetry of the filament.

\subsubsection{Results of modal analysis}
Having shown that only four eigenvalues are non-zero, and derived a method by which to calculate them analytically, we now present these results, and exploit them to calculate the swimming speed. Denoting the two positive eigenvalues by $\lambda_+$ and $\lambda_-$, with corresponding eigenfunctions $g_+$ and $g_-$ respectively, Eq.~\eqref{eq:sum} simplifies to
\begin{equation}
\begin{split}
\mathcal{U} = \left[\left(\int_0^1 f(s)g_+(s)~ds\right)^2 - \left(\int_0^1 f(s)g_+(1-s)~ds\right)^2\right]\lambda_+& \\
+ \left[\left(\int_0^1 f(s)g_-(s)~ds\right)^2 - \left(\int_0^1 f(s)g_-(1-s)~ds\right)^2\right]\lambda_-&.
\end{split}
\label{eq:modal1}
\end{equation}
In particular, if the artificial filament is powered by $M$ discrete actuators, modelled as $\delta$-functions with strengths $F_m$ and positions $\xi_m$, this formula can be written as
\begin{equation}
\begin{split}
\mathcal{U} = \left[\left(\sum_{m=1}^M F_kg_+(\xi_k)\right)^2 - \left(\sum_{m=1}^M F_kg_+(1-\xi_k)\right)^2\right]\lambda_+& \\
+ \left[\left(\sum_{m=1}^M F_kg_-(\xi_k)\right)^2 - \left(\sum_{m=1}^M F_kg_-(1-\xi_k)\right)^2\right]\lambda_-&.
\end{split}
\label{eq:modal2}
\end{equation}

\begin{figure}
\centering
\includegraphics[width = \linewidth]{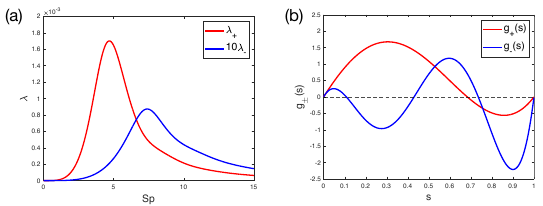}
\caption{{\bf Eigenvalues and eigenfunctions for monophasic forcing.} (\textbf{a}) The large (red) and small (blue) eigenvalues, $\lambda_+$ and $\lambda_-$ respectively, for monophasic forcing, $\phi \equiv 0$. {Note that $\lambda_+$ is plotted against $10\lambda_-$.} (\textbf{b}) Eigenfunctions corresponding to the large (red) and small (blue) eigenvalues, $g_+$ and $g_-$ respectively, normalised with the established fixed forcing magnitude condition, for the optimal sperm number $Sp = 4.7$.
}
\label{Figure_3}
\end{figure}

These equations yield significant results, enabling $\mathcal{U}$ to be determined through only four sums or integrals. Again, this represents a reduction in complexity from $\mathcal{O}(N^2)$ to $\mathcal{O}(N)$ ($N$ discretisation points for a continuous $f$) or from $\mathcal{O}(M^2)$ to $\mathcal{O}(M)$ (number $M$ of discrete actuators) compared to Eq.~\eqref{eq:U_G_swim}. However, whilst useful for computation of $\mathcal{U}$ for a particular $f$ and $Sp$, these equations are not, \textit{a priori}, conducive to analysis, nor to any optimisation over $f$ and $Sp$ besides brute-force methods, and thus require further simplification.

 \subsubsection{Finding the optimal sperm number and neglecting the smaller eigenvalue}

We find that both eigenvalues have a maximum at a single value of $Sp$, as illustrated in Fig.~\ref{Figure_3}a (Note that, since $\lambda_+$ is much larger than $\lambda_-$, we plot $\lambda_+$ against $10\lambda_-$). We see that $\lambda_+$ has a maximum of $\lambda_+ \approx 0.00170$ at $Sp \approx 4.70$, where it is around a hundred times larger than $\lambda_-$ at the same point. Furthermore, the maximum value of $\lambda_+$ is around 20 times larger than that of $\lambda_-$. Since $\lambda_+$ is much larger than $\lambda_-$, we set the sperm number to be $Sp = 4.7$, {for the remainder of this example}. For this optimal $Sp$, the eigenfunctions $g_+$ and $g_-$ (for comparison) are shown in Fig.~\ref{Figure_3}b. 

Since $\lambda_+ \gg \lambda_-$, an approximate solution may be obtained by neglecting in the analysis the smaller eigenvalue and eigenfunction. The swimming speed under continuous forcing $f$ can then be expressed approximately as
\begin{equation}\label{eq:332}
\mathcal{U} \approx \left[\left(\int_0^1 f(s)g_+(s)~ds\right)^2 - \left(\int_0^1 f(s)g_+(1-s)~ds\right)^2\right]\lambda_+,
\end{equation}
or, for $M$ discrete actuators,
\begin{equation}\label{eq:333}
\mathcal{U} \approx \left[\left(\sum_{m=1}^M F_kg_+(\xi_k)\right)^2 - \left(\sum_{m=1}^M F_kg_+(1-\xi_k)\right)^2\right]\lambda_+.
\end{equation}

These formulae represent further improvement. Since we have set $Sp$ to take its approximate optimal value, optimising the swimming speed over choice of $f$ therefore produces the approximate global maximum value of $\mathcal{U}$ across all values of $Sp$, eliminating the need to vary $Sp$ manually. These formulae also require only half as many calculations compared to Eqs.~\eqref{eq:modal1} and~\eqref{eq:modal2}. 

\subsubsection{Swimming speed as the difference of two squares}

\begin{figure}
\centering
\includegraphics[width = \linewidth]{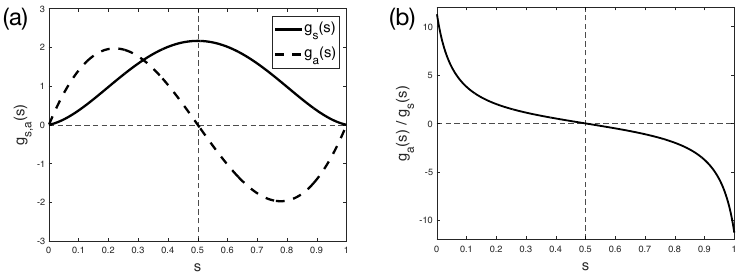}
\caption{{\bf Symmetric and antisymmetric decomposition of eigenmodes.} (\textbf{a}) Symmetric  function $g_s$ (thick solid line) and antisymmetric function $g_a$ (thick dashed line); the lines $g = 0$ and $s = 0.5$ are shown as thin dashed lines. (\textbf{b}) Ratio $g_a(s)/g_s(s)$ used for optimisations.}
\label{Figure_4}
\end{figure}

Aiming to find the simplest mathematical method to maximise the value of $\mathcal{U}$, we now define {symmetric and antisymmetric functions, $g_s$ and $g_a$ respectively, to be twice the symmetric and antisymmetric }components of the eigenfunction $g_+$,
\begin{equation}
\begin{split}
g_s(s) &= g_+(s) + g_+(1-s), \\
g_a(s) &= g_+(s) - g_+(1-s).
\end{split}
\label{eq:g_s_a}
\end{equation}
These two functions are shown in Fig.~\ref{Figure_4}a, with their ratio (to be used later) plotted in Fig.~\ref{Figure_4}b. The swimming speed from Eq.~\eqref{eq:332}
 is then given by
\begin{equation}
\mathcal{U} \approx \left[\int_0^1 f(s)g_s(s)~ds\right]\left[\int_0^1 f(s)g_a(s)~ds\right]\lambda_+,
\label{eq:int_speed}
\end{equation}
or for discrete actuators, Eq.~\eqref{eq:333} 
becomes 
\begin{equation}
\mathcal{U} \approx \left[\sum_{m=1}^M F_mg_s(\xi_m)\right]\left[\sum_{m=1}^M F_mg_a(\xi_m)\right]\lambda_+.
\label{eq:sum_speed}
\end{equation}
In optimising the filament, we may assume (without loss of generality) that both integrals/sums are non-negative.

\subsubsection{Application to  artificial swimmers with continuous forcing}

An artificial swimmer is unlikely to be constrained by the fixed forcing magnitude condition given by Eq.~\eqref{eq:ffmc}; instead the limitations are more likely to be in the engineering and fabrication of the forcing on such small scales. We will apply a more appropriate constraint below, but for now let us consider the simple situation where each location $\xi$ along the filament is either passive ($f(\xi) = 0$) or forced ($f(\xi) = 1$). This is a {reasonable parallel to} the artificial microswimmer demonstrated in Ref.~\cite{Dreyfus_et_al_2005}, where forcing is provided by magnetic beads (where $f(\xi) \neq 0$){, potentially} alternating with inert sections of filament ($f(\xi) = 0$). {Note however that this specific example~\cite{Dreyfus_et_al_2005} utilises externally powered actuation, and the resultant dynamics cannot be exactly described by the model developed in this paper. Nonetheless, it is an intuitive example of the form of a microswimmer that utilises piecewise constant forcing.} By using Eq.~\eqref{eq:int_speed}, we now determine the choice of $f$ which maximises the swimming speed.

\begin{figure}
\centering
\includegraphics[width = \linewidth]{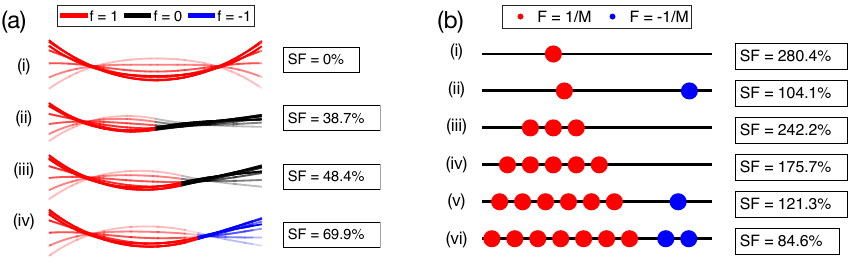}
\caption{{\bf Artificial swimmers with continuous {(a) and discrete (b)} forcing.}  (\textbf{a}) Example artificial swimmers, {with continuous, piecewise constant forcing, }of progressively increasing speeds, optimised using Eq.~\eqref{eq:int_speed}. Solid lines represent the filaments at $t = 0$, and $t$ increases to $\pi$ for progressively fading lines. Speeds have been divided by $\lambda_+$ to produce the speed factor (SF) as a percentage. {(\textbf{i}) Uniformly forced filament that cannot swim due to the scallop theorem. (\textbf{ii}) Simple swimmer with forcing in the front half only. (\textbf{iii}) Swimmer with frontal forcing occupying an optimal fraction of the filament. (\textbf{iv}) Swimmer with optimally chosen positive and negative forcing.} 
{An animated version of these swimmers is shown in  Supplementary Video 1.}(\textbf{b}) Example artificial swimmers for $M = 1, 2, 3, 5, 7, 9$ discrete actuators, each of strength $1/M$, with speed factors indicated. Minimum spacing of $0.1$ between actuators. {An animated version of these swimmers is shown in Supplementary Video 2.}}
\label{Figure_5}
\end{figure}

Despite the temptation of using as much forcing as possible, simply setting $f = 1$ everywhere  cannot produce any swimming due to symmetry of the filament (Fig.~\ref{Figure_5}a.i), and a more detailed analysis is required. Using Eq.~\eqref{eq:int_speed}, with the functions $g_s$ and $g_a$ given in Fig.~\ref{Figure_4}a, we see that, regardless of the values taken by $f(\xi)$ for $\xi > 0.5$, the swimming speed is always increased by setting $f(\xi) = 1$ for all $\xi \le 0.5$ (assuming  {without loss of generality} that both integrals in Eq.~\eqref{eq:int_speed} are non-negative). We immediately deduce that a possible swimmer  has $f(\xi) = 1$ for $\xi \le 0.5$ and $f(\xi) = 0$ for $\xi > 0.5$ (Fig.~\ref{Figure_5}a.ii).

We can further improve this swimmer by setting $f(\xi) = 1$ for some values of $\xi > 0.5$. Suppose that $f$ takes its  optimal value. Then any acceptable change in $f$ (i.e.~any change which maintains $f(\xi) = 0$ or $f(\xi) = 1$ at each $\xi$) will necessarily reduce the swimming speed.  {Calling such a change $\delta f(s; \xi)$, which takes a constant non-zero value (either $1$ or $-1$) in a small region of width $\Delta$ around $s = \xi$, and is zero elsewhere, the swimming speed is perturbed by
\begin{equation}
\delta\mathcal{U} \approx \left\{\left[\int_0^1 f(s)g_s(s)~ds\right]g_a(\xi) + g_s(\xi)\left[\int_0^1 f(s)g_a(s)~ds\right]\right\}\Delta \lambda_+\delta f(\xi; \xi).
\end{equation}}
{For an optimal $f$, we require this to be negative whenever $\delta f$ takes an allowable value. Therefore, if $f(\xi) = 1$, then $\delta \mathcal{U}$ must be negative for $\delta f(\xi; \xi) = -1$, whilst if $f(\xi) = 0$, then $\delta \mathcal{U}$ must be negative for $\delta f(\xi; \xi) = 1$. }We deduce that
\begin{align}
g_a(\xi)\left(\int_0^1 f(s)g_s(s)~ds\right) + g_s(\xi)\left(\int_0^1 f(s)g_a(s)~ds\right) &\ge 0 ~~~~~~~~~~ {\rm when}~~f(\xi) = 1, \\
g_a(\xi)\left(\int_0^1 f(s)g_s(s)~ds\right) + g_s(\xi)\left(\int_0^1 f(s)g_a(s)~ds\right) &\le 0 ~~~~~~~~~~  {\rm when}~~f(\xi) = 0.
\end{align}
By noting that $g_s$ is positive, we can write these inequalities as
\begin{align}
\frac{\int_0^1 f(s)g_a(s)~ds}{\int_0^1 f(s)g_s(s)~ds} + \frac{g_a(\xi)}{g_s(\xi)} &\ge 0 ~~~~~~~~~~  {\rm when}~~f(\xi) = 1, \\
\frac{\int_0^1 f(s)g_a(s)~ds}{\int_0^1 f(s)g_s(s)~ds} + \frac{g_a(\xi)}{g_s(\xi)} &\le 0 ~~~~~~~~~~  {\rm when}~~f(\xi) = 0.
\end{align}
As shown in Fig.~\ref{Figure_4}b, the ratio  $g_a/g_s$ is an antisymmetric, strictly decreasing function of $\xi$. Therefore, assuming that $f$ does not take the same constant value for all $\xi$, an optimal $f$ that satisfies these inequalities must have a corresponding point $\xi_1^*$ such that
\begin{equation}
\frac{\int_0^1 f(s)g_a(s)~ds}{\int_0^1 f(s)g_s(s)~ds} = -\frac{g_a(\xi_1^*)}{g_s(\xi_1^*)},
\end{equation}
with $f(\xi) = 1$ for $\xi < \xi_1^*$ and $f(\xi) = 0$ for $\xi > \xi_1^*$. Hence this implicit equation can be simplified as
\begin{equation}\label{eq:342}
\frac{\int_0^{\xi_1^*} g_a(s)~ds}{\int_0^{\xi_1^*} g_s(s)~ds} = -\frac{g_a(\xi_1^*)}{g_s(\xi_1^*)}.
\end{equation}
We can solve Eq.~\eqref{eq:342}   numerically to identify $\xi_1^* = 0.625$, with the corresponding swimmer illustrated in Fig.~\ref{Figure_5}a.iii.

This procedure can easily be adapted to consider a somewhat more advanced swimmer where $f$ can be $1$, $0$ or $-1$. Following the same logic we obtain
\begin{align}
\frac{\int_0^1 f(s)g_a(s)~ds}{\int_0^1 f(s)g_s(s)~ds} + \frac{g_a(\xi)}{g_s(\xi)} &\ge 0 ~~~~~~~~~~ {\rm when}~~f(\xi) = 1, \\
\frac{\int_0^1 f(s)g_a(s)~ds}{\int_0^1 f(s)g_s(s)~ds} + \frac{g_a(\xi)}{g_s(\xi)} &= 0 ~~~~~~~~~~ {\rm when}~~f(\xi) = 0 \label{eq:345},\\
\frac{\int_0^1 f(s)g_a(s)~ds}{\int_0^1 f(s)g_s(s)~ds} + \frac{g_a(\xi)}{g_s(\xi)} &\le 0 ~~~~~~~~~~ {\rm when}~~f(\xi) = -1.
\end{align}
 Once again, there is therefore a point $\xi_2^*$ for which $f(\xi) = 1$ for $\xi < \xi_2^*$ and $f(\xi) = -1$ for $\xi > \xi_2^*$.  {There can be no region where $f = 0$ because that would require Eq.~\eqref{eq:345} to be satisfied within said region, which is impossible for the strictly decreasing ratio $g_a/g_s$. Then} $\xi_2^*$ is now given by
\begin{equation}\label{eq:346}
\frac{\int_0^{\xi_2^*} g_a(s)~ds - \int_{\xi_2^*}^1 g_a(s)~ds}{\int_0^{\xi_2^*} g_s(s)~ds - \int_{\xi_2^*}^1 g_a(s)~ds} = -\frac{g_a(\xi_2^*)}{g_s(\xi_2^*)}.
\end{equation}
The implicit equation in Eq.~\eqref{eq:346} can be solved to find $\xi_2^* = 0.701$, resulting in the swimmer  shown in Fig.~\ref{Figure_5}a.iv.  

The  four artificial swimmers shown in Fig.~\ref{Figure_5}a have progressively increasing speeds, as indicated in insets and quantified as speed fraction (SF), defined as the dimensionless speed divided by {the} eigenvalue $\lambda_+$ and  interpreted as the fraction of the swimming speed $\mathcal{U}$ that {would be} achieved by setting the forcing function $f$ to be the eigenfunction $g_+$. We see in particular that this final artificial swimmer can achieve a maximum of $69.9\%$ of its optimal speed{, as defined by setting $f = g_+$. Animated versions of the swimmers of Fig.~\ref{Figure_5}a are available in Supplementary Video 1.}

\subsubsection{Piecewise constant forcing is optimised in the limit of single-point forcing}

An optimal speed fraction of $69.9\%$ (Fig.~\ref{Figure_5}a.iv) is {fairly respectable}, and an artificial swimmer with this piecewise constant forcing would be {a comparably fast swimmer to} one that uses eigenfunction forcing. It is notable that  the swimmer in Fig.~\ref{Figure_5}a.iv has a swimming speed that is less than double the swimming speed of the swimmer of Fig.~\ref{Figure_5}a.ii, despite having twice as much total forcing. {Indeed, uniformly doubling the forcing magnitude (from $f(\xi)$ to $2f(\xi)$) of any given swimmer would quadruple the swimming speed, which makes an improvement of less than double   disappointing,} and this motivates a new forcing constraint. An artificial swimmer is likely to be limited by the engineering involved in its fabrication, and this is likely to manifest as a total forcing magnitude constraint of the form
\begin{equation}
\int_0^1 |f(\xi)|~d\xi = 1.
\label{eq:ffmc2}
\end{equation}

A specific example of this would be the artificial {millimeter-scale }swimmer powered by cardiomyocytes (heart muscle cells) studied experimentally in  Ref.~\cite{Williams_Anand_Rajagopalan_Saif_2014}, where each cardiomyocyte generates a fixed forcing moment. In that case, the most likely limitation is the number of muscle cells that can be cultured onto the filament, corresponding to the new forcing magnitude constraint in Eq.~\eqref{eq:ffmc2}. {Note that, under this constraint, $g_+$ has a total forcing magnitude of approximately $0.84$, rather than $1$. Eigenfunction forcing under this new total forcing magnitude constraint (i.e. $f = g_+/0.84$) will therefore produce a speed factor of $142.0\%$.}

Returning to Fig.~\ref{Figure_5}a, we observe that a swimmer that has $f(\xi) = 2$ for $\xi \le 0.5$ and $f(\xi) = 0$ for $\xi > 0.5$ (i.e.~as in Fig.~\ref{Figure_5}a.ii but with doubled forcing and hence quadrupled speed) would swim more than twice as fast as one with the more distributed forcing $|f(\xi)| = 1$ for all $\xi$ (Fig.~\ref{Figure_5}a.iv), for the same total forcing magnitude as defined by Eq.~\eqref{eq:ffmc2}. This suggests that distributing the forcing along a greater length is, perhaps counterintuitively, detrimental to swimming. This is in contrast to the biological situation of spermatozoa, which use distributed forcing along the entire axoneme~\cite{Machin_1958, Rikmenspoel_1965, Brokaw_1970, Rikmenspoel_1966}. 

In fact, as we are now going to show, in this new constraint swimming is optimised by taking the limit of single-point actuation; this will turn out to  present up to four-fold improvement compared to the swimmer of Fig.~\ref{Figure_5}a.iv, whilst still satisfying Eq.~\eqref{eq:ffmc2}.

From previous swimmers illustrated in Fig.~\ref{Figure_5}a, it is apparent from  the signs of $g_s$ and $g_a$ (Fig.~\ref{Figure_4}a) and of their ratio $g_a/g_s$ (Fig.~\ref{Figure_4}b), that positive forcing ($f \ge 0$) should exist in $0 \le \xi \le 0.5$, and extending some way past $0.5$, whilst negative forcing ($f \le 0$) should occupy the remainder of the filament.

In the range $0 \le \xi \le 0.221$, both $g_s$ and $g_a$ are increasing (Fig.~\ref{Figure_4}a) and so any positive forcing in this range should be moved right if possible.  Conversely, for $0.5 \le \xi \le 0.779$, both $g_s$ and $g_a$ are decreasing and any positive forcing should be moved to the left. In the region $0.221 \le \xi \le 0.5$, it is easily shown that $g_s$ and $g_a$ are both concave functions, and thus contracting any positive forcing (e.g.~changing $f(\xi) = F$ for $\xi_1 \le \xi \le \xi_2$ into $f(\xi) = F\times \left(\xi_2 - \xi_1\right)/\left(\xi_2 - \xi_1 - 2\epsilon\right)$ for $\xi_1 + \epsilon \le \xi \le \xi_2 - \epsilon$) will necessarily increase the swimming speed, since it would be increasing both terms in Eq.~\eqref{eq:int_speed}.  From these observations, we deduce that the benefit to the swimming speed contributed by the positive forcing can be improved by contracting it, ideally to the limit of single-point actuation at some position in the interval $0.221 \le \xi \le 0.5$. 

Conversely, any negative forcing in the range $0.5 \le \xi \le 0.779$ should be moved rightwards; in particular, this implies that all negative forcing should exist entirely to the right of any positive forcing. While $g_s$ is slightly concave in a small sub-region at the start of $0.779 \le \xi \le 1$, it is convex in most of the region, with $g_a$ being convex in the entire region. Overall, swimming speed is therefore essentially increased by a contraction of the negative forcing in this region.  Therefore we deduce that negative forcing should again be contracted as much as possible, ideally to the limit of single-point actuation at some point in the range $0.779 \le \xi \le 1$, to produce  near-optimal swimming speed.

\subsubsection{Optimising  swimmers under discrete   actuation}

Having shown that piecewise constant forcing is optimised in the limit of single-point forcing, we now investigate numerically the optimal configurations{, directly applying the results to the biohybrid swimmer of Ref.~\cite{Williams_Anand_Rajagopalan_Saif_2014} shortly}.  Starting with the case of a single-point positive forcing, $f(\xi) = \delta(\xi - \xi_1)$, and no negative forcing, we find that swimming speed is maximised by placing the actuator at $\xi_1 = 0.309$, achieving a far greater swimming speed fraction of $280.4\%$ (see swimmer illustrated in Fig.~\ref{Figure_5}b.i). {It is perhaps surprising that concentrating all the forcing in a single location can produce a far greater swimming speed compared even to eigenfunction forcing under the same fixed forcing magnitude constraint (Eq.~\eqref{eq:ffmc2}), with the speed factor almost doubling.} Meanwhile, suppose we instead want to use one positive and one negative forcing actuator of equal strength, $1/2$. We find these should be placed at $\xi_1 = 0.357$ and $\xi_2 = 0.902$ respectively, though this only achieves a speed fraction of $104.1\%$ (Fig.~\ref{Figure_5}b.ii). 

By varying over all possible allocations of forcing magnitudes between these two actuators, we find that just having a single, positive actuator of strength $1$ at $\xi_1 = 0.309$, and no negative actuator (i.e.~the swimmer of Fig.~\ref{Figure_5}b.i), is optimal among all possible discrete   actuations.  Across the entire suite of forcing functions, the optimal swimming speed subject to the new constraint Eq.~\eqref{eq:ffmc2} is thus achieved using this single-point actuator.

From a practical standpoint, there may be situations in which an artificial swimmer cannot be constructed with all of its forcing applied to one location. For example, in the cardiomyocytes-powered swimmer  from Ref.~\cite{Williams_Anand_Rajagopalan_Saif_2014}, the magnitude of the forcing that can be applied at a particular location is limited by the contractile force of the cardiomyocytes that power the filament. Since the filament has limited space and one cannot simply place arbitrarily many of these cardiomyocytes at $\xi_1 = 0.309$, the forcing would likely  need to be distributed over some length, using some number $M$ of single-point actuators, possibly limited by some minimum spacing requirement between the actuators.

If the locations of the actuators were pre-determined (e.g.~$\xi_m = (2m-1)/2M$), and we are free to choose the forcing strengths ($F_m = 1/M$, $0$ or $-1/M$), then the results in Fig.~\ref{Figure_5}a serve as an excellent guide when $M$ is large. Meanwhile, brute force computations are feasible when $M$ is small, since there are only $3^M$ configurations and their number can be further reduced using the logic we have considered above (e.g.~noting that actuators in $\xi \le 0.5$ should have positive forcing).

On the other hand, the actuator locations may not be pre-defined, and we may instead be free to choose the optimal actuator location and forcing direction for a given number of actuators. As we have  argued above, all actuators with the same forcing should be placed as close together as possible.  We may assume (by symmetry) that at least half of the actuators have positive forcing, giving at most $M/2$ configurations for the number of actuators of each sign. 
For each of these, we are left with just two continuous parameters to optimise: the locations $\xi_1$ and $\xi_2$ of the centre of each group of actuators. In Fig.~\ref{Figure_5}b (iii - vi) we illustrate the results obtained for  $M = 3, 5, 7, 9$,  assuming   a minimum actuator separation of $0.1$. As expected, we   see that distributing the forcing over multiple actuators progressively decreases the swimming speed. For larger values of $M$, the inclusion of negative forcing actuators becomes necessary to achieve optimal swimming speed (Fig.~\ref{Figure_5}b.v - vi). {Animated versions of the swimmers of Fig.~\ref{Figure_5}b are available in Supplementary Video 2.}

{\subsubsection{Application to the biohybrid swimmer of Ref.~\cite{Williams_Anand_Rajagopalan_Saif_2014}}

The results of this paper, in particular those of Fig.~\ref{Figure_5}b, can be applied to perform optimisations based on various constraints such as actuator separation and direction. A good example of such a swimmer is the approximately $2~$mm long biohybrid swimmer produced in Ref.~\cite{Williams_Anand_Rajagopalan_Saif_2014}. This swimmer (which, in the experiments, possessed a passive head) was actuated by cardiomyocytes, cultured onto the flagellum near the point of attachment with the head. By applying our model, we can identify the optimal configuration for a headless swimmer, and compare our results to the numerical results obtained in Ref.~\cite{Williams_Anand_Rajagopalan_Saif_2014} in the limit of a vanishing head. Simply applying the same value of $Sp$ ($4.19$) and the same actuator location (around $\xi = 0.25$) used in their experiments, we calculate a dimensional swimming speed of around $1.3~\mu$m/s, in good agreement with their numerical results ($1$ to $2~\mu$m/s). This agreement is despite the fact that the dimensionless moment forcing has a magnitude of around $4.5$, suggesting our model remains reasonably accurate even for non-linear actuation. Furthermore, the cardiomyocyte contractions were not simple sinusoidal functions of time, and higher temporal modes were present, though this is easily resolved since the temporal modes decouple and the overall dimensional swimming is simply the sum of the dimensional swimming speeds corresponding to each temporal mode. Modifying $Sp$ (for example, by elongating the swimmer) and $\xi$ to take their optimal values of $4.7$ and $0.309$ respectively (see Fig.~\ref{Figure_5}b.i) can increase this dimensional swimming speed, measured in swimmer lengths per second, by around $50~$\%. Importantly, this is an immediate result requiring no additional analysis or computation.}

\section{Conclusion}
\label{S:4}

The dynamics of slender filaments is relevant not only for modelling biological microorganisms such as spermatozoa, but also for the design and fabrication of artificial swimmers. By revisiting and linearising classical elastohydrodynamic theory, we have derived simple expressions for the shape of the filament (Eq.~\eqref{eq:Psi_expression}) in terms of the Green's function ({Appendix \ref{appendix:G}}) and for the swimming speed (Eq.~\eqref{eq:U_G_swim}) in terms of the swimming speed function (Eq.~\eqref{eq:G_swim}). In particular, the swimming speed may be evaluated, for a given forcing, without the need to explicitly calculate the shape of the filament, in an $\mathcal{O}\left(N^2\right)$ process{, where $N$ is the number of points used to perform the numerical integration in Eq.~\eqref{eq:U_G_swim}}. However, while useful for the quick and easy evaluation of the swimming speed, this process is not conducive to optimisations over choices of forcing functions.

{The real symmetric nature of the swimming speed function suggests a modal approach, and} we next numerically identified its eigenfunctions and eigenvalues by discretising the swimming speed function as a large symmetric matrix. We further revealed via calculus of variations that these eigenfunctions provide optimal forcing functions for swimming, assuming a fixed forcing magnitude constraint (Eq.~\eqref{eq:ffmc}). These eigenfunctions and eigenvalues provide an alternative method by which to calculate the swimming speed (Eq.~\eqref{eq:sum}), and in particular most of the modes may be neglected by virtue of small eigenvalues. We have found that, in a variety of situations, only a small number of eigenmodes need be retained to produce accurate results (Fig.~\ref{Figure_2}), thereby reducing the computational complexity in evaluating the swimming speed to an $\mathcal{O}(N)$ problem{, or even an $\mathcal{O}(1)$ problem if the forcing function is initially constructed in terms of the modal basis.}


Furthermore, this modal approach allows for optimisations of the forcing function, such as by approximating it as a sum of the dominant modes. In the particular situation of monophasic forcing, the eigenfunctions and eigenvalues may be calculated analytically, and in particular only four of the infinitely many eigenvalues are non-zero. These consist of two symmetric pairs, one of which is dominated by the other at optimal $Sp$ (Fig.~\ref{Figure_3}). By neglecting the lesser eigenvalue, we have revealed a wide range of analyses that allow for the optimal design of an artificial swimmer, subject to various constraints (Fig.~\ref{Figure_5}), in particular producing effective swimmers that utilise {piecewise constant forcing or} single-point actuation, relevant to previously demonstrated artificial swimmers~\cite{Dreyfus_et_al_2005, Williams_Anand_Rajagopalan_Saif_2014}.

It is clear that the slender filaments considered are but a  simplified model system. Real swimmers, both biological and artificial, typically carry a body. Not only does this allow for the transport of a payload, but the presence of a body can also be advantageous for producing increased swimming speed ~\cite{Williams_Anand_Rajagopalan_Saif_2014}. Furthermore, we have considered here only two-dimensional filament motion, whilst some spermatozoa can exhibit three-dimensional (often helical) motion~\cite{Brokaw_1966, Ishijima_Sekiguchi_Hiramoto_1988, Woolley_Vernon_2001}. {Finally, our assumption of small ($\epsilon \ll 1$) disturbances will inevitably lead to inaccuracies when modelling real swimmers with $\mathcal{O}(1)$ disturbances, though previous investigations have shown remarkable agreement even in such situations~\cite{Yu_Lauga_Hosoi_2006}.} Despite these simplifications, we have obtained results which are quantitatively similar to those observed in spermatozoa swimming through \textit{in vitro} fertilisation medium (Fig.~\ref{Figure_2}). Potential adaptations and improvements to this work include {generalisations of the mathematical approach, such as considering different variational constraints which may produce new and interesting optimisations. Future work could also consider generalisations of the swimmer design and behaviour, such as through the inclusion of a head, as in spermatozoa and the aforementioned artificial swimmers, which have an effective head either in the form of a payload~\cite{Dreyfus_et_al_2005}, or a designed head that improves swimming~\cite{Williams_Anand_Rajagopalan_Saif_2014}.}

We hope that this work will be helpful in the design and optimisation of artificial swimmers, and in aiding the ongoing understanding of the motion of spermatozoa and other microorganisms, with potential applications to fertility science, micro-engineering and general medical applications.
 
\appendix 
\renewcommand\theequation{\thesection~\arabic{equation}}

\section{Green's function for the hyperdiffusion equation} \label{appendix:G}
The Green's function $G(s; \xi)$ used in Eq.~\eqref{eq:Psi_expression} is the solution to the equation
\begin{equation}
Sp^4iG + G'''' = \delta(s - \xi).
\label{eq:G}
\end{equation}
Note this corresponds to a single point actuator forcing located at position $\xi$ along the filament. To enforce the zero force and moment boundary conditions on the filament, we require $G = G_s = 0$ at $s = 0$ and $s = 1$. Meanwhile, as standard for a Green's function, $G$, $G'$ and $G''$ must be continuous at $s = \xi$, with $G'''$ jumping by a value of $1$. The solution for $G$ is given by
\begin{equation}
G = H(s - \xi)G_+ + H(\xi - s)G_-,
\end{equation}
where
\begin{align}
G_-(s) &= A_-e^{Sp\eta s} + B_-e^{Sp\eta i s} + C_-e^{-Sp \eta s} + D_-e^{-Sp \eta i s} \label{G_minus},\\
G_+(s) &= A_+e^{Sp\eta s} + B_+e^{Sp\eta i s} + C_+e^{-Sp \eta s} + D_+e^{-Sp \eta i s} \label{G_plus},
\end{align}
with $\eta = e^{-\pi i/8}$. {The coefficients are determined by the boundary and jump conditions, via the matrix equation
\begin{equation}\tiny
\left[\begin{matrix} 1 & 1 & 1 & 1 & 0 & 0 & 0 & 0 \\ 1 & i & -1 & -i & 0 & 0 & 0 & 0 \\ 0 & 0 & 0 & 0 & e^{Sp\eta} & e^{Sp\eta i} & e^{-Sp \eta} & e^{-Sp\eta i} \\ 0 & 0 & 0 & 0 & e^{Sp\eta} & ie^{Sp\eta i} & -e^{-Sp \eta} & -ie^{-Sp\eta i}\\-e^{Sp\eta \xi} & -e^{Sp\eta i \xi} & -e^{-Sp \eta \xi} & -e^{-Sp\eta i \xi} & e^{Sp\eta \xi} & e^{Sp\eta i \xi} & e^{-Sp \eta \xi} & e^{-Sp\eta i \xi} \\ -e^{Sp\eta \xi} & -ie^{Sp\eta i \xi} & e^{-Sp \eta \xi} & ie^{-Sp\eta i \xi} & e^{Sp\eta \xi} & ie^{Sp\eta i \xi} & -e^{-Sp \eta \xi} & -ie^{-Sp\eta i \xi} \\ -e^{Sp\eta \xi} & e^{Sp\eta i \xi} & -e^{-Sp \eta \xi} & e^{-Sp\eta i \xi} & e^{Sp\eta \xi} & -e^{Sp\eta i \xi} & e^{-Sp \eta \xi} & -e^{-Sp\eta i \xi} \\ -e^{Sp\eta \xi} & ie^{Sp\eta i \xi} & e^{-Sp \eta \xi} & -ie^{-Sp\eta i \xi} & e^{Sp\eta \xi} & -ie^{Sp\eta i \xi} & -e^{-Sp \eta \xi} & ie^{-Sp\eta i \xi}\end{matrix}\right]\left[\begin{matrix}A_- \\ B_- \\ C_- \\ D_-\\ A_+ \\ B_+ \\ C_+ \\ D_+ \end{matrix}\right] = \left[\begin{matrix}0 \\ 0 \\ 0 \\ 0 \\ 0 \\ 0 \\ 0 \\ \frac{1}{Sp^3\eta^3} \end{matrix}\right].
\label{eq:matrix_equation}
\end{equation}}
We have therefore solved the hyperdiffusion equation for the Green's function $G$.

\section{Derivation of Eq.~\eqref{eq:U_general}}\label{appendix:U}
Recalling that
\begin{equation}
\mathcal{U} = -2\int^1_0\left<\Psi_{sss}\Psi_{sst}\right> ds,
\end{equation}
we derive here Eq.~\eqref{eq:U_general} for $\mathcal{U}$, for general moment forcing $m^{(1)}(s, t) = \Re\left[f(s)e^{-i\phi(s)}e^{it}\right]$ and corresponding solution $\Psi(s, t) = \Re\left[e^{it}\int_0^1 G(s; \xi)f(\xi)e^{-i\phi(\xi)}~d\xi\right]$. We begin by substituting the solution for $\Psi$, moving the time-average inside the integrals:
\begin{equation}
\mathcal{U} = -2\int_{s = 0}^1 \int_{\xi_1 = 0}^1 \int_{\xi_2 = 0}^1 \left<\Re\left[e^{it} G'''(s; \xi_1)f(\xi_1)e^{-i\phi(\xi_1)}\right] \Re\left[ie^{it} G''(s; \xi_2)f(\xi_2)e^{-i\phi(\xi_2)}\right]\right>~d\xi_2d\xi_1~ds.
\end{equation}
Next we evaluate the time-averages to find
\begin{equation}
\begin{split}
\mathcal{U} = \int_{s = 0}^1 \int_{\xi_1 = 0}^1 \int_{\xi_2 = 0}^1 &\left\{ -\Im\left[G'''(s; \xi_1)f(\xi_1)e^{-i\phi(\xi_1)}\right] \Re\left[G''(s; \xi_2)f(\xi_2)e^{-i\phi(\xi_2)}\right] \right. \\
&~~~\left. + \Re\left[G'''(s; \xi_1)f(\xi_1)e^{-i\phi(\xi_1)}\right] \Im\left[G''(s; \xi_2)f(\xi_2)e^{-i\phi(\xi_2)}\right] \right\} ~d\xi_2d\xi_1ds.
\end{split}
\end{equation}
We integrate by parts with respect to $s$, recalling that $G'(0; \xi) = G'(1; \xi) = 0$,
\begin{equation}
\begin{split}
\mathcal{U} = \int_{s = 0}^1 \int_{\xi_1 = 0}^1 \int_{\xi_2 = 0}^1 &\left\{\Im\left[G''''(s; \xi_1)f(\xi_1)e^{-i\phi(\xi_1)}\right] \Re\left[G'(s; \xi_2)f(\xi_2)e^{-i\phi(\xi_2)}\right] \right.\\
&\left. - \Re\left[G''''(s; \xi_1)f(\xi_1)e^{-i\phi(\xi_1)}\right] \Im\left[G'(s; \xi_2)f(\xi_2)e^{-i\phi(\xi_2)}\right] \right\}~d\xi_2d\xi_1ds.
\end{split}
\end{equation}
We now recall the governing equation for the Green's function, $Sp^4iG + G'''' = \delta(s - \xi)$, giving
\begin{equation}
\begin{split}
\mathcal{U} = \int_{s = 0}^1 \int_{\xi_1 = 0}^1 \int_{\xi_2 = 0}^1 &\left\{ \Im\left[\left(\delta(s - \xi_1) - Sp^4iG(s; \xi_1)\right)f(\xi_1)e^{-i\phi(\xi_1)}\right] \Re\left[G'(s; \xi_2)f(\xi_2)e^{-i\phi(\xi_2)}\right] \right.\\
&\left. - \Re\left[\left(\delta(s - \xi_1) - Sp^4iG(s; \xi_1)\right)f(\xi_1)e^{-i\phi(\xi_1)}\right] \Im\left[G'(s; \xi_2)f(\xi_2)e^{-i\phi(\xi_2)}\right] \right\}~d\xi_2d\xi_1ds.
\end{split}
\end{equation}
Using integration by parts with respect to $s$, recalling that $G(0; \xi) = G(1; \xi) = 0$, we note that
\begin{equation}
\begin{split}
I(\xi_1, \xi_2) &= \int_{s = 0}^1 \Re\left[G(s; \xi_1)f(\xi_1)e^{-i\phi(\xi_1)}\right]\Re\left[G'(s; \xi_2)f(\xi_2)e^{-i\phi(\xi_2)}\right]~ds \\
&= -\int_{s = 0}^1 \Re\left[G'(s; \xi_1)f(\xi_1)e^{-i\phi(\xi_1)}\right]\Re\left[G(s; \xi_2)f(\xi_2)e^{-i\phi(\xi_2)}\right]~ds = -I(\xi_2, \xi_1).
\end{split}
\end{equation}
Therefore
\begin{equation}
\int_{\xi_1 = 0}^1 \int_{\xi_2 = 0}^1 I(\xi_1, \xi_2)~d\xi_2 d\xi_1
= \int_{\xi_1 = 0}^1 \int_{\xi_2 = 0}^1 -I(\xi_2, \xi_1)~d\xi_2 d\xi_1
= \int_{\xi_2 = 0}^1 \int_{\xi_1 = 0}^1 -I(\xi_1, \xi_2)~d\xi_1 d\xi_2,
\end{equation}
and so is zero. The same argument holds for $I(\xi_1, \xi_2)$ defined using imaginary parts rather than real parts. Therefore the equation for $\mathcal{U}$ simplifies to
\begin{equation}
\begin{split}
\mathcal{U} = \int_{s = 0}^1 \int_{\xi_1 = 0}^1 \int_{\xi_2 = 0}^1 &\left\{\Im\left[\delta(s - \xi_1)f(\xi_1)e^{-i\phi(\xi_1)}\right] \Re\left[G'(s; \xi_2)f(\xi_2)e^{-i\phi(\xi_2)}\right]\right. \\
&\left.- \Re\left[\delta(s - \xi_1)f(\xi_1)e^{-i\phi(\xi_1)}\right] \Im\left[G'(s; \xi_2)f(\xi_2)e^{-i\phi(\xi_2)}\right]\right\}~d\xi_2d\xi_1ds.
\end{split}
\end{equation}
The $\delta$-functions are real and so
\begin{equation}
\begin{split}
\mathcal{U} = \int_{s = 0}^1 \int_{\xi_1 = 0}^1 \int_{\xi_2 = 0}^1 &\left\{\delta(s - \xi_1)\Im\left[f(\xi_1)e^{-i\phi(\xi_1)}\right] \Re\left[G'(s; \xi_2)f(\xi_2)e^{-i\phi(\xi_2)}\right]\right. \\
&\left. - \delta(s - \xi_1)\Re\left[f(\xi_1)e^{-i\phi(\xi_1)}\right] \Im\left[G'(s; \xi_2)f(\xi_2)e^{-i\phi(\xi_2)}\right]\right\}~d\xi_2d\xi_1ds,
\end{split}
\end{equation}
and we now evaluate the integral in $s$ to obtain
\begin{equation}
\begin{split}
\mathcal{U} = \int_{\xi_1 = 0}^1 \int_{\xi_2 = 0}^1 &\left\{\Im\left[f(\xi_1)e^{-i\phi(\xi_1)}\right] \Re\left[G'(\xi_1; \xi_2)f(\xi_2)e^{-i\phi(\xi_2)}\right]\right. \\
&\left. - \Re\left[f(\xi_1)e^{-i\phi(\xi_1)}\right] \Im\left[G'(\xi_1; \xi_2)f(\xi_2)e^{-i\phi(\xi_2)}\right] \right\}~d\xi_2d\xi_1.
\end{split}
\end{equation}
Recalling $f$ is real gives
\begin{equation}
\mathcal{U} = \int_{\xi_1 = 0}^1 \int_{\xi_2 = 0}^1 f(\xi_1)\left(\Im\left[e^{-i\phi(\xi_1)}\right] \Re\left[G'(\xi_1; \xi_2)e^{-i\phi(\xi_2)}\right] - \Re\left[e^{-i\phi(\xi_1)}\right] \Im\left[G'(\xi_1; \xi_2)e^{-i\phi(\xi_2)}\right]\right)f(\xi_2)~d\xi_2d\xi_1.
\end{equation}
And by re-combining terms, we obtain
\begin{equation}
\mathcal{U} = -\int_{\xi_1 = 0}^1 \int_{\xi_2 = 0}^1 f(\xi_1)\Im\left[G'(\xi_1; \xi_2)e^{i(\phi(\xi_1) - \phi(\xi_2))}\right]f(\xi_2)~d\xi_2d\xi_1,
\end{equation}
as promised.

\section{Standard eigenfunction calculation for monophasic forcing} \label{appendix:eig}
The method presented in the main text allows for the concise calculation of the eigenfunctions and eigenvalues for the case $\phi \equiv 0$. We present here a more standard method that has potential to be generalised to different situations, at the cost of being more cumbersome and producing an $8 \times 8$ system with four zero eigenvalues.

Suppose $f(\xi_1) = \Re\left[e^{k\xi_1}\right] = \frac{1}{2}\left(e^{k\xi_1} + e^{k^*\xi_1}\right)$, where $k$ is one of the four values such that this is a natural mode; $k^4 = -Sp^4 i$. Note that $k^*$ does not correspond to a natural mode. Then we can solve for $I_1$ and $I_2$ as
\begin{align}
I_1^{(1, k)}(\xi_1) &=  \frac{\xi_1 e^{k\xi_1}}{8k^3} + \frac{e^{k^*\xi_1}}{4Sp^4 i} + A_1^{(1, k)}e^{Sp\eta \xi_1} + B_1^{(1, k)}e^{Sp\eta i \xi_1} + C_1^{(1, k)}e^{-Sp\eta \xi_1} + D_1^{(1, k)}e^{-Sp\eta i \xi_1}, \\
I_2^{(1, k)}(\xi_1) &=  \frac{\xi_1 e^{k\xi_1}}{8k^2} + \frac{k^*e^{k^*\xi_1}}{4Sp^4 i} + A_2^{(1, k)}e^{Sp\eta \xi_1} + B_2^{(1, k)}e^{Sp\eta i \xi_1} + C_2^{(1, k)}e^{-Sp\eta \xi_1} + D_2^{(1, k)}e^{-Sp\eta i \xi_1}.
\end{align}
Alternatively, if $f(\xi_1) = \Re\left[ie^{k\xi_1}\right]$, then $I_1$ and $I_2$ are given by
\begin{align}
I_1^{(i, k)}(\xi_1) &=  \frac{i\xi_1 e^{k\xi_1}}{8k^3} - \frac{e^{k^*\xi_1}}{4Sp^4} + A_1^{(i, k)}e^{Sp\eta \xi_1} + B_1^{(i, k)}e^{Sp\eta i \xi_1} + C_1^{(i, k)}e^{-Sp\eta \xi_1} + D_1^{(i, k)}e^{-Sp\eta i \xi_1}, \\
I_2^{(i, k)}(\xi_1) &=  \frac{i\xi_1 e^{k\xi_1}}{8k^2} - \frac{k^*e^{k^*\xi_1}}{4Sp^4} + A_2^{(i, k)}e^{Sp\eta \xi_1} + B_2^{(i, k)}e^{Sp\eta i \xi_1} + C_2^{(i, k)}e^{-Sp\eta \xi_1} + D_2^{(i, k)}e^{-Sp\eta i \xi_1}.
\end{align}
The 64 coefficients $A, B, C, D$ can easily be calculated, for a particular $Sp$, according to the boundary conditions $I(0) = I(1) = I'(0) = I'(1) = 0$. From these, we can define, for $b = 1$ or $b = i$, and a particular value of $k$,
\begin{equation}
I_3^{(b, k)} = \frac{1}{2}i\left(\left(I_1^{(b, k)}\right)' - I_2^{(b, k)}\right),
\end{equation}
which, defining new coefficients,
\begin{equation}
\begin{split}
A_3^{(b, k)} &= \frac{1}{2}i\left(Sp\eta A_1^{(b, k)} - A_2^{(b, k)}\right), \\
B_3^{(b, k)} &= \frac{1}{2}i\left(Sp\eta iB_1^{(b, k)} - B_2^{(b, k)}\right), \\
C_3^{(b, k)} &= \frac{1}{2}i\left(-Sp\eta C_1^{(b, k)} - C_2^{(b, k)}\right), \\
D_3^{(b, k)} &= \frac{1}{2}i\left(-Sp\eta iD_1^{(b, k)} - D_2^{(b, k)}\right),
\end{split}
\end{equation}
can be evaluated as
\begin{equation}
I_3^{(b, k)}(\xi_1) = A_3^{(b, k)}e^{Sp\eta \xi_1} + B_3^{(b, k)}e^{Sp\eta i \xi_1} + C_3^{(b, k)}e^{-Sp\eta \xi_1} + D_3^{(b, k)}e^{-Sp\eta i \xi_1}.
\end{equation}
Importantly, the 32 coefficients can be evaluated easily for any given $Sp$. Therefore, letting general $f(\xi_1)$ be given by the sum of its eight possible terms,
\begin{equation}
f(\xi_1) = \sum_{\substack{b = 1, i \\ k^4 = -Sp^4 i}}E^{(b, k)}\Re\left[ae^{k\xi_1}\right],
\end{equation}
where each $E^{(b, k)}$ is real, we obtain an overall $I_3$ given by
\begin{equation}
I_3(\xi_1) = \sum_{\substack{b = 1, i \\ k^4 = -Sp^4 i}}  E^{(b, k)}\left(A_3^{(b, k)}e^{Sp\eta \xi_1} + B_3^{(b, k)}e^{Sp\eta i \xi_1} + C_3^{(b, k)}e^{-Sp\eta \xi_1} + D_3^{(b, k)}e^{-Sp\eta i \xi_1}\right).
\end{equation}
The eigenvalue problem can then be written as
\begin{equation}
\begin{split}
&\lambda\sum_{\substack{b = 1, i \\ k^4 = -Sp^4 i}}E^{(b, k)}\Re\left[be^{k\xi_1}\right] \\
= &\sum_{\substack{b = 1, i \\ k^4 = -Sp^4 i}}  E^{(b, k)}\Re\left[A_3^{(b, k)}e^{Sp\eta \xi_1} + B_3^{(b, k)}e^{Sp\eta i \xi_1} + C_3^{(b, k)}e^{-Sp\eta \xi_1} + D_3^{(b, k)}e^{-Sp\eta i \xi_1}\right].
\end{split}
\end{equation}
Equating each of the eight modes gives an $8 \times 8$ eigenvalue problem, for a vector $\mathbf{v}$ of the coefficients $E^{(b, k)}$, and matrix $\mathbf{M}_{eig}$,
\begin{equation}
\lambda \mathbf{v} = \mathbf{M}_{eig} \mathbf{v}.
\end{equation}
The easiest way to write this explicitly is to define the \textit{vectorisation} of a set of coefficients by
\begin{equation}
\mathbf{vec(F)} = \left[\begin{matrix}F^{(1, Sp\eta)} \ F^{(1, Sp\eta i)} \ F^{(1, -Sp\eta)} \ F^{(1, -Sp\eta i)} \ F^{(i, Sp\eta)} \ F^{(i, Sp\eta i)} \ F^{(i, -Sp\eta)} \ F^{(i, -Sp\eta i)} \end{matrix}\right].
\end{equation}
Then
\begin{equation}
\mathbf{v} = \mathbf{vec(E)}^\intercal,
\end{equation}
and
\begin{equation}
\mathbf{M}_{eig} = \left[\begin{matrix}\Re\left[\mathbf{vec(A_3)}\right] \\ \Re\left[\mathbf{vec(B_3)}\right] \\ \Re\left[\mathbf{vec(C_3)}\right] \\ \Re\left[\mathbf{vec(D_3)}\right] \\ \Im\left[\mathbf{vec(A_3)}\right] \\ \Im\left[\mathbf{vec(B_3)}\right] \\ \Im\left[\mathbf{vec(C_3)}\right] \\ \Im\left[\mathbf{vec(D_3)}\right]\end{matrix}\right].
\end{equation}
As mentioned previously, four of the eight resultant eigenvalues are zero, leaving four non-zero eigenvalues, as promised.


\bibliography{bibliography}
\bibliographystyle{RS}

\end{document}